\documentclass[onecolumn]{aastex631}
\usepackage{amsmath, amsthm, amssymb, amsfonts}
\usepackage{multirow}
\usepackage{graphicx}
\usepackage{fontawesome}

\newcommand{\summ}{\sum_{l=0}^{L}}
\newcommand{\Aint}{A_{n,\text{int}}}

\usepackage[margins]{trackchanges}
\addeditor{none} 
\addeditor{CR}
\addeditor{NB}

\shorttitle{Spherical Harmonics for 1D Radiative Transfer II}
\shortauthors{Rooney et al.}

\begin{document}

\title{Spherical Harmonics for the 1D Radiative Transfer Equation II: Thermal Emission}

\correspondingauthor{Natasha E. Batalha}
\email{natasha.e.batalha@nasa.gov}

\author[0000-0001-9005-2872]{Caoimhe M. Rooney}
\affiliation{NASA Ames Research Center,
Moffett Field,
CA, 94035 USA}

\author[0000-0003-1240-6844]{Natasha E. Batalha}
\affiliation{NASA Ames Research Center,
Moffett Field, 
CA, 94035 USA}

\author[0000-0002-5251-2943]{Mark S. Marley}
\affiliation{Department of Planetary Sciences, Lunar and Planetary Laboratory, University of Arizona, Tucson AZ 85721}

\begin{abstract}

Approximate methods to estimate solutions to the radiative transfer equation are essential for the understanding of atmospheres of exoplanets and brown dwarfs. 
The simplest and most popular choice is the ``two-stream method'' which is often used to produce simple yet effective models for radiative transfer in scattering and absorbing media.
\cite{toon1989rapid} (Toon89) outlined a two-stream method for computing reflected light and thermal spectra and was later implemented in the open-source radiative transfer model \texttt{PICASO}.
In Part~I of this series, we developed an analytical spherical harmonics method for solving the radiative transfer equation for reflected solar radiation \citep{rooney2023spherical} which was implemented in \texttt{PICASO} to increase the accuracy of the code by offering a higher-order approximation.
This work is an extension of this spherical harmonics derivation to study thermal emission spectroscopy.
We highlight the model differences in the approach for thermal emission and benchmark the 4-term method (SH4) against Toon89 and a high-stream discrete-ordinates method, \texttt{CDISORT}.  
By comparing the spectra produced by each model we demonstrate that the SH4 method provides a significant increase in accuracy, compared to Toon89, which can be attributed to the increased order of approximation and to the choice of phase function.
We also explore the trade-off between computational time and model accuracy. We find that our 4-term method is twice as slow as our 2-term method, but is up to five times more accurate, when compared with \texttt{CDISORT}. Therefore, SH4 provides excellent improvement in model accuracy with minimal sacrifice in numerical expense.

\end{abstract}



\section{Introduction} \label{sec:intro}
Studying the atmospheres of planets and substellar objects relies on computationally efficient methods to solve the radiative transfer equation in scattering and absorbing media.
However, exact solutions typically do not exist.
Scientists rely on approximate, parameterized methods to estimate solutions, but inclusion of intricate microphysical detail can render these models computationally intractable for useful applications \citep{stephens1984multimode, thomas2002radiative, chandrasekhar1960radiative, liou2002introduction}.
The goal of radiative transfer parameterization in numerical models for exoplanet and brown dwarf atmospheres is to provide computationally efficient yet accurate methods to calculate radiative fluxes and heating rates \citep{stephens1984parameterization}. 
The approach to solving the radiative transfer equation is frequently chosen through a balance between accuracy and computational efficiency. 
In many practical cases, there are significant uncertainties associated with defining characteristics of the atmosphere, such as the composition, scattering phase function and opacities.
Such uncertainties often dominate over small model errors in the solution, and therefore, obtaining a computationally efficient solution often takes precedence.

The most popular approximate methods for solving the radiative transfer equation are the (1) discrete-ordinates method \citep{chandrasekhar1960radiative, stamnes1988numerically, stamnes2000disort}, (2) Monte-Carlo method \citep{modest2013radiative, iwabuchi2006efficient} and (3) spherical harmonics method \citep{modest1989modified, modest2013radiative, olfe1967modification, https://doi.org/10.48550/arxiv.2205.09713}. 
The general approach of the discrete-ordinates method (DOM) is to discretize the solid angle by a finite number $L$ of directions or ``streams'', along which the radiative intensities are tracked.
DISORT \citep{stamnes1988numerically, stamnes2000disort} is an example of a discrete ordinate algorithm for radiative transfer that is capable of simulating thermal emission, absorption, and scattering for arbitrary phase functions across the electromagnetic spectrum.
The convergence of DOM can depreciate for optically thick media \citep{modest2013radiative, fiveland1996acceleration, lewis1984computational}, however, there exist a number of acceleration schemes to improve the convergence rate of DOM \citep{fiveland1996acceleration, lewis1984computational}.
Monte-Carlo methods track emitted photons throughout the media, and although accepted to be a largely accurate method, it is computationally taxing which makes it unsuitable for some applications \citep{iwabuchi2006efficient, mayer2009radiative}. 
The spherical harmonics (SH) approximation, denoted $P_{L-1}$, operates by expanding the intensity and phase function into a series of $L$ spherical harmonics, or Legendre polynomials. This decouples spatial and directional dependencies.
This method involves fewer equations than DOM and is potentially more accurate with comparable computational expense, but higher order expansions are mathematically complex and increasingly difficult to implement as $L$ increases \citep{ge2015implementation, https://doi.org/10.48550/arxiv.2205.09713}. 

Such models have been studied for reflected solar radiation \citep[e.g., most recently in our Part~1.][]{rooney2023spherical}.
For example, the two-stream discrete-ordinates ($L=2$) and two-term spherical harmonics ($P_1$) techniques are often used to provide simple yet effective models for atmospheric radiative transfer and are widely considered some of the simplest and most prolific approximations \citep{meador1980two, king1986comparative, chandrasekhar1960radiative, mihalas2013foundations, https://doi.org/10.48550/arxiv.2205.09713, li1996four, zhang2013doubling}.
Two-stream methods are most useful for obtaining angle-averaged quantities such as heating rates and albedos \citep{schuster1905radiation, meador1980two, heng2018radiative}. 

These studies of reflected solar radiation have shown that even though the two-stream methods are computationally preferable, they are often unsuitable for certain physical conditions.
For example, non-physical solutions to the two-stream method are obtained for the case of a collimated incident beam \citep{meador1980two}.
However, there exist model adjustments to correct for these limitations in the two-stream method.
In particular, applying the delta ($\delta$)-adjustment to the two-stream technique improves the accuracy of radiative flux calculations by taking into account strong forward scattering due to large particles \citep{joseph1976delta, wiscombe1977delta, king1986comparative, liou1988simple}.

Though these corrections often help to improve accuracy, it has also been shown that improved accuracy can be achieved by considering four-stream approximations \citep{liou1974analytic, liou1988simple, cuzzi1982delta, rooney2023spherical}. Four-stream methods are also still able to leverage the $\delta$-approximation to allow for strong forward scattering.
However, an increase in the order of approximation comes with the penalty of an increase in computational expense.
Four-stream approximations are significantly more efficacious in cases of non-isotropic scattering and can even be used in general circulation models \citep{liou1988simple, heng2018radiative}. Their performance has been explored for both homogeneous and inhomogeneous atmospheres in reflected solar radiation \citep{liou1973numerical, liou1988simple, fu1991parameterization, shibata1992accuracy}.

As well as solar radiation, these approximations can also be applied to study scattering in the presence of thermal emission \citep{mihalas1978stellar, toon1989rapid, fu1997multiple}, which is the focus of this work. 
Infrared scattering is essential to understand emission spectra of exoplanets and brown dwarfs \citep{taylor2021does}, due to the ubiquity of clouds in atmospheres \citep{marley2013,gao2021universal} and the defining role they play in sculpting thermal emission. Specifically, the thermal emission for some classes of extrasolar planets and brown dwarfs (e.g., the L dwarfs) arises, in some wavelengths, from within the scattering, absorbing cloud layers. Therefore, in those cases it is particularly important to treat the radiative transfer within the cloud as carefully as possible. Additionally, the radiative-equilibrium temperature structure of an atmosphere \citep[e.g.,][]{Mukherjee2023picaso} depends upon the difference between upwards and downwards incident and emergent fluxes. Overall, an accurate treatment within the scattering cloud decks is required to have confidence in computed thermal profiles. 

Two-stream methods have been implemented for infrared thermal radiation, such as the well-known work of
\cite{toon1989rapid}, who derived a general two-stream solution for the upward and downward fluxes within a single homogeneous layer. 
By considering continuity of flux across a number of stacked, homogeneous layers, the single-layer solution is extended to a multi-layer atmosphere.
The final solution is obtained through the two-stream source function technique, with the source function written in terms of the two-stream intensity. As a side note, related to the two-stream technique are popular analytical calculations of temperature-pressure profiles to further understand the thermal atmospheric structure of atmospheres \citep{hubeny2003possible, hansen2008target, guillot2010radiative, heng2012stability, heng2015atmospheric, robinson2012analytic, parmentier2014non}.

The \cite{toon1989rapid} methodology has in particular been utilized extensively for the study of planetary and substellar atmospheres \citep[e.g.][]{mckay1989,marley1999,burrows1997,fortney2005,marley2021}  and this implementation is  available
in open-source Python code \texttt{PICASO} \citep{natasha_batalha_2022_6419943}. 
This approach, however, is currently limited to two-stream approximations.
Despite its usefulness in radiative transfer calculations due to its simplicity and ease of implementation, \cite{toon1989rapid} reported that relative errors in the emissivity calculated by such approaches can be as much as 10\% in optically thin cases. 

In addition to the potential errors reported by \cite{toon1989rapid}, it has also been shown that increasing the approximations to four-streams improves the accuracy of the infrared models, similar to the reflected solar case \citep{fu1997multiple, liou1988simple, lin2013upward}.
\cite{fu1997multiple} applied the $\delta$-two and four-stream discrete-ordinates method \citep{chandrasekhar1960radiative, stamnes1988numerically, stamnes2000disort} to solve the infrared radiative transfer equation in a vertically inhomogeneous atmosphere.
By comparing the approximations to the high-order $\delta$-129-stream model, 
the authors found that the $\delta$-two-stream scheme can produce acceptable results under most atmospheric conditions, but suffers from large errors for small optical depth.
The $\delta$-four-stream method yields high accuracy in radiative fluxes and heating rates under all atmospheric conditions considered, however, the authors acknowledge a significant increase in computational cost.
\cite{zhang2016analytical} also investigated $\delta$-two and four-stream discrete-ordinates for infrared radiative transfer, demonstrating an analytical approach.
By comparing the methods to a $\delta$-64-stream DOM method, the authors similarly conclude that the four-stream method outperforms two-streams, particularly for small optical depths, reporting relative errors as high as 15\% for two-stream versus 2\% for four-stream. 

These studies motivated the development of an analytical spherical harmonics method for solving the radiative transfer equation. The first component for solar radiation was published recently in \cite{rooney2023spherical}.  
In a similar vein to the \cite{toon1989rapid} methodology, \cite{rooney2023spherical} derived and solved a system of equations for the upwards and downwards fluxes at every layer of our atmosphere, with the critical difference of a $\delta$-adjusted four-term spherical harmonics ($P_3$) approximation in place of Toon's two-stream approach.
By applying the source-function technique to calculate the azimuthally averaged intensity emerging from the top of a vertically inhomogeneous atmosphere, we can compute the spectrum for clear and cloudy planets or brown dwarfs. Though the spherical harmonics approaches for reflected light and thermal emission are largely identical, the primary differences lie in the source terms, boundary conditions and the related applications. 

Therefore, the present work is an extension to the derivation in \cite{rooney2023spherical}, namely, applying the spherical harmonics model to thermal emission spectroscopy. We aim to make this manuscript easily cross-referenced with the numerical method implemented within \texttt{PICASO} source code. Throughout the manuscript, we refer the reader to \cite{rooney2023spherical} for more intricate detail into the derivation of the spherical harmonics method for 1D radiative transfer, when necessary. Here, we include only the key mathematical expressions that define the thermal emission model.
We have also included persistent hyperlinks that can be accessed by clicking the following icon: \href{https://natashabatalha.github.io/picaso/}{\faCode}, that will redirect the reader to the relevant lines of code (stored on Github) corresponding to the relevant mathematical expression.

We outline this work as follows: in Sections \ref{sec:SH} and \ref{sec:source_function}, we briefly explain the derivation of the spherical harmonics (SH) method for thermal radiation.
As aforementioned, the SH method for reflected light and thermal emission are largely identical, with the exception of the source term and boundary conditions.
In this section, we focus on the differences incurred by considering the thermal source term and relevant boundary conditions.
We consider both two and four-stream approximations for plane-parallel atmospheres of many layers, where we apply the source-function technique to handle the multi-layer aspect of the model.

In Section \ref{sec:analysis}, we compare the two and four-term spherical harmonics models and the \cite{toon1989rapid} approach implemented in \texttt{PICASO} to a 16 and 32-stream discrete ordinates method, \texttt{CDISORT} to illustrate the accuracy gains by increasing the number of streams from two to four.
We also explore the impact of this order increase on computational time, and discuss the timing-accuracy trade-off that might be considered when choosing a model.

\section{Solving the radiative transfer equation using spherical harmonics}
\label{sec:SH}
We wish to use the spherical harmonics technique to solve the azimuthally-averaged, one-dimensional radiative transfer equation:
\begin{equation}
	\mu\frac{\partial I}{\partial \tau}(\tau,\mu) = I(\tau,\mu) 
		- \frac{w_0}{2}\int_{-1}^{1} I(\tau,\mu')P(\mu,\mu')\mathrm{d}\mu'
		- 2\pi(1-w_0)B(T),
	\label{eq:RTE}
\end{equation}
where the location within the atmosphere is specified by $\tau\in[0,\tau_N]$, (where $\tau_N$ is the cumulative optical depth), $I$ is the azimuthally averaged intensity and $w_0$ is the single scattering albedo, $B(T)$ is the Planck function at temperature $T$, and $P(\mu,\mu')$ is the azimuthally averaged scattering phase function.

We note the similarities between the radiative transfer equation for thermal emission \eqref{eq:RTE} and that for reflected light, outlined in \cite{rooney2023spherical}.
The difference lies in the final term on the right-hand side, the source term $S(T)$, defined as
\begin{align}
    S(T) = 
    \begin{cases}
        2\pi(1-w_0)B(T), &\qquad \text{thermal emission},\\
        \frac{w_0}{4 \pi}F_\odot e^{-\frac{\tau}{\mu_0}}P(\mu,-\mu_0), &\qquad \text{reflected light}.
    \end{cases}
    \label{eq:source_terms}
\end{align}
We emphasise that all other terms in the azimuthally-averaged, one-dimensional radiative transfer equation \eqref{eq:RTE} are identical for reflected light and thermal emission.
This allows us to largely follow the spherical harmonics model derivation outlined in \cite{rooney2023spherical} for reflected light, with a few modifications to allow for the different source term. 
We will highlight these differences throughout this work. However, we refer the reader to \cite{rooney2023spherical} for a more in-depth discussion of the general model derivation.

By expanding the phase function and intensity in terms of Legendre polynomials $P_l$, up to given order $L$:
\begin{align}
	P(\mu,\mu') &= \sum_{l=0}^L \chi_l P_l(\mu) P_l(\mu'),\label{eq:PF}\\
	I(\tau,\mu) &= \sum_{l=0}^L (2l+1) I_l(\tau) P_l(\mu),\label{eq:I}
\end{align}
where the coefficients $\chi_l$ of the phase function expansion can be determined from the orthogonal property of Legendre polynomials \citep{liou2002introduction}:
\begin{equation}
    \chi_l = \frac{2l+1}{2}\int_{-1}^1 P(\cos\Theta)P_l(\cos\Theta)\mathrm{d}\cos\Theta.
    \label{eq:chi_def}
\end{equation}
we can substitute \eqref{eq:PF} and \eqref{eq:I} into \eqref{eq:RTE} and use both the orthogonality property and recursion relation of Legendre polynomials to obtain
\begin{equation}
	\sum_{l=0}^L \left[(l+1)\frac{\mathrm{d}I_{l+1}}{\mathrm{d}\tau}
			+ l\frac{\mathrm{d}I_{l-1}}{\mathrm{d}\tau}\right]P_l(\mu)
            = \sum_{l=0}^L [a_lI_l(\tau)  
			- b_l\delta_{0l}]
			P_l(\mu).
	\label{eq:RTE_expanded}
\end{equation}
Here, $\delta_{0l}$ is the Dirac-delta function ($\delta_{0l}=1$ for $l=0$, and 0 otherwise) and
\begin{align}
    a_l &= (2l+1)-w_0\chi_l,\label{eq:a_l}\\
    b_l &= 2\pi(1-w_0)B(T),\label{eq:b_l}
\end{align}
for $l=0,\cdots,L$.
Here, $a_l$ is identical to that derived for reflected light \citep{rooney2023spherical}, whereas the expressions for $b_l$ differ.
This is because the source term \eqref{eq:source_terms} is relevant only for the $b_l$ terms.
Thus, any analysis involving only the $a_l$ terms and not the $b_l$ terms will be identical for reflected light and thermal emission.

We assume that the Planck function with a single layer $B(T)$ can be represented as a Taylor series expansion \citep[as done in][]{toon1989rapid}, namely
\begin{equation}
    B(T(\tau)) = B_0 + B_1\tau,
\end{equation}
where $B_0$ is the Planck function evaluated at $\tau=0$ (or the top of the layer) and $B_1$ is related to the Planck function at temperature $T_\text{bot}$ at the bottom of the layer $\tau_N$:
\begin{equation}
    B_1 = \frac{B(T_\text{bot}) - B_0}{\tau_N}
\end{equation}

\subsection{$P_1$ Multiple Layers}
For clarity and to demonstrate the spherical harmonics methodology, \cite{rooney2023spherical} began with an atmosphere consisting of a single horizontally homogeneous layer, before extending the analysis to the more practical case of multiple layers.
Here, we proceed immediately to the multiple-layer solution.

Let us first study the two-stream spherical harmonics problem, denoted $P_1$, where $L=1$ represents the highest Legendre polynomial in the expansion.
Consider an atmosphere consisting of $N$ horizontally homogeneous layers, where layer $n$ is characterized by single scattering albedo $w_{0,n}$, asymmetry parameter $g_{0,n}$ and optical thickness $\partial\tau_n = \tau_n-\tau_{n-1}$ for $n=1,\dots,N$.

To solve the radiative transfer equation \eqref{eq:RTE} in the $n^\text{th}$ layer we rescale the optical depth as
\begin{equation}
	\hat\tau = \tau - \tau_{n-1},\qquad \hat\tau\in[0,\partial\tau_n].
\end{equation}
Dropping the hats, we continue with the solutions within layer $n$ for $\tau\in[0,\partial\tau_n]$.

We can formulate \eqref{eq:RTE_expanded}
as a matrix system within layer $n$:
\begin{equation}
	\frac{\mathrm{d}}{\mathrm{d}\tau}
	\begin{pmatrix}
		I_{0,n} \\ I_{1,n}
	\end{pmatrix} = 
	\begin{pmatrix}
		0 & a_{1,n} \\ a_{0,n} & 0
	\end{pmatrix}  
	\begin{pmatrix}
		I_{0,n} \\ I_{1,n}
	\end{pmatrix} - 
	\begin{pmatrix}
		b_1 \\ b_0
	\end{pmatrix},
	\label{eq:P1_sys}
\end{equation}

Closely following the methodology outlined in \cite{rooney2023spherical}, we arrive at the layer-wise solution:
\begin{equation}
	\begin{pmatrix}
		I_{0,n} \\ I_{1,n}
	\end{pmatrix} = 
	\begin{pmatrix}
		e^{-\lambda_n\tau} & e^{\lambda_n\tau} \\ 
		-q_n e^{-\lambda_n\tau} & q_n e^{\lambda_n\tau} 
	\end{pmatrix}  
	\begin{pmatrix}
		X_{0,n} \\ X_{1,n}
	\end{pmatrix} 
	+\frac{2\pi(1-w_{0,n})}{a_{0,n}} 
	\begin{pmatrix}
		B_{0,n}+\tau B_{1,n} \\ \frac{B_{1,n}}{a_1}
	\end{pmatrix},
	\label{eq:P1_nlayer_sol}
\end{equation}
for $\tau\in[0,\partial\tau_n]$, where
\begin{align}
    a_{l,n} &= (2l+1)-w_{0,n}\chi_{l,n},
\end{align}
is the multi-layer extension of \eqref{eq:a_l}, and 
\begin{equation}
    \lambda_n=\sqrt{a_{0,n} a_{1,n}}, \qquad q_n=\lambda_n/a_{1,n}.
\end{equation}
Following \cite{rooney2023spherical}, we can rewrite system \eqref{eq:P1_nlayer_sol} in terms of fluxes,
\begin{equation}
	\begin{pmatrix}
		F_n^- \\ F_n^+
	\end{pmatrix} = 
	\begin{pmatrix}
		Q_n^+ e^{-\lambda_n\tau} & Q_n^- e^{\lambda_n\tau} \\ 
		Q_n^- e^{-\lambda_n\tau} & Q_n^+ e^{\lambda_n\tau} 
	\end{pmatrix}  
	\begin{pmatrix}
		X_{0,n} \\ X_{1,n}
	\end{pmatrix} + 
	\begin{pmatrix}
		Z_n^- \\ Z_n^+
	\end{pmatrix},
	\label{eq:P1_nlayer_flx}
\end{equation}
where $Q_n^\pm=\pi(1\pm 2q_n)$ \href{https://github.com/natashabatalha/picaso/blob/9d4cbd672a75c1faf5297c3f1d74074018cd7ef3/picaso/fluxes.py#L3102-L3103}{\faCode} and $Z_n^{\pm}$ given by \href{https://github.com/natashabatalha/picaso/blob/9d4cbd672a75c1faf5297c3f1d74074018cd7ef3/picaso/fluxes.py#L3117-L3120}{\faCode} 
\begin{equation}
    Z_n^\pm(\tau) = \frac{\pi(1-w_{0,n})}{a_{0,n}}\left(B_{1,n}\tau + B_{0,n} \pm \frac{2}{a_{1,n}}B_{1,n}\right),
\end{equation}
with boundary conditions \href{https://github.com/natashabatalha/picaso/blob/9d4cbd672a75c1faf5297c3f1d74074018cd7ef3/picaso/fluxes.py#L3130-L3139}{\faCode} 
\begin{align}
	F_1^-(0) &= 0,\\
	F_n^-(\partial\tau_n) &= F_{n+1}^-(0),\\
	F_n^+(\partial\tau_n) &= F_{n+1}^+(0),
\end{align}
and
\begin{align}
    F_N^+(\tau_N) =\begin{cases} \pi\left(B(\tau_N) + \frac{2}{3}\frac{\partial B}{\partial\tau}(\tau_N)\right), &\quad \text{non-terrestrial}, \\
    \pi B(\tau_N) +A_SF^-(\tau_N)&\quad \text{terrestrial, hard surface},
    \end{cases}
    \label{eq:bottom_flx_BC_P1}
\end{align}
where $A_S$ is the surface reflectivity.
The final boundary condition \eqref{eq:bottom_flx_BC_P1} is derived from \cite{mihalas1978stellar} in Appendix 
\ref{app:f_BC}.
These boundary conditions enforce that there is no incident diffuse flux at the top of the atmosphere, and the upward flux at the surface is either that from a (potentially) reflective surface or an estimate of the upwards flux emerging from an atmosphere that continues below the lowermost model grid point (e.g., a giant planet or brown dwarf atmosphere).
The spherical harmonics flux problem is formulated in \texttt{PICASO} by representing the system in terms of banded matrices, and solved using the \texttt{solve\_banded} functionality of \texttt{SciPy} \citep{2020SciPy-NMeth} \href{https://github.com/natashabatalha/picaso/blob/891343fcc41faa345f8b85aaa8d50c4939c421a3/picaso/fluxes.py#L3532}{\faCode}.

\subsection{P3 Multiple Layers}
\label{sec:P3}
Similarly, the $P_3$ problem for multiple layers has the solution:
\begin{equation}
	\begin{pmatrix}
		I_{0,n} \\ I_{1,n} \\ I_{2,n} \\ I_{3,n}
	\end{pmatrix} = 
	\begin{pmatrix}
		e^{-\lambda_{1,n}\tau} & e^{\lambda_{1,n}\tau} & 
			e^{-\lambda_{2,n}\tau} & e^{\lambda_{2,n}\tau} \\ 
		R_{1,n} e^{-\lambda_{1,n}\tau} & -R_{1,n} e^{\lambda_{1,n}\tau} & 
			R_{2,n} e^{-\lambda_{2,n}\tau} & -R_{2,n} e^{\lambda_{2,n}\tau} \\ 
		Q_{1,n} e^{-\lambda_{1,n}\tau} & Q_{1,n} e^{\lambda_{1,n}\tau} & 
			Q_{2,n} e^{-\lambda_{2,n}\tau} & Q_{2,n} e^{\lambda_{2,n}\tau} \\ 
		S_{1,n} e^{-\lambda_{1,n}\tau} & -S_{1,n} e^{\lambda_{1,n}\tau} & 
			S_{2,n} e^{-\lambda_{2,n}\tau} & -S_{2,n} e^{\lambda_{2,n}\tau} 
	\end{pmatrix}  
	\begin{pmatrix}
		X_{0,n} \\ X_{1,n} \\ X_{2,n} \\ X_{3,n}
	\end{pmatrix} - 
	\frac{2\pi(1-w_{0,n})}{a_{0,n}}
	\begin{pmatrix}
		B_{0,n}+\tau B_{1,n} \\ 
		\frac{B_{1,n}}{a_{1,n}} \\ 
		0 \\ 0
	\end{pmatrix},
	\label{eq:P3_nlayer_sol}
\end{equation}
for $\tau\in[0,\partial\tau_n]$, 
where \href{https://github.com/natashabatalha/picaso/blob/9d4cbd672a75c1faf5297c3f1d74074018cd7ef3/picaso/fluxes.py#L3238-L3241}{\faCode}
\begin{align}
    \lambda_{1,2,n} = \sqrt{\frac{1}{2}(\beta_n \pm \sqrt{\beta_n^2-4\gamma_n})},\qquad  \beta_n=a_{0,n}a_{1,n} + \frac{1}{9}a_{2,n}a_{3,n} + \frac{4}{9}a_{0,n}a_{3,n},\qquad \gamma_n=\frac{1}{9}a_{0,n}a_{1,n}a_{2,n}a_{3,n},
\end{align}
and \href{https://github.com/natashabatalha/picaso/blob/9d4cbd672a75c1faf5297c3f1d74074018cd7ef3/picaso/fluxes.py#L3273-L3275}{\faCode}
\begin{align}
    R_{1,2,n}=-\frac{a_{0,n}}{\lambda_{1,2,n}},\qquad
    Q_{1,2,n} = \frac{1}{2}\left(\frac{a_{0,n}a_{1,n}}{\lambda_{1,2,n}^2} - 1\right),\qquad
    S_{1,2,n} = -\frac{3}{2a_{3,n}}\left(\frac{a_{0,n}a_{1,n}}{\lambda_{1,2,n}} - \lambda_{1,2,n}\right).
    \label{eq:RnQnSn}
\end{align}

This problem can be written in terms of fluxes as
\begin{equation}
	\begin{pmatrix}
		F_n^- \\ f_n^- \\ F_n^+ \\ f_n^+
	\end{pmatrix} = 
	\begin{pmatrix}
		p_{1,n}^-e^{-\lambda_{1,n}\tau} & p_{1,n}^+e^{\lambda_{1,n}\tau} & 
			p_{2,n}^-e^{-\lambda_{2,n}\tau} & p_{2,n}^+e^{\lambda_{2,n}\tau} \\ 
		q_{1,n}^-e^{-\lambda_{1,n}\tau} & q_{1,n}^+e^{\lambda_{1,n}\tau} & 
			q_{2,n}^-e^{-\lambda_{2,n}\tau} & q_{2,n}^+e^{\lambda_{2,n}\tau} \\ 
		p_{1,n}^+e^{-\lambda_{1,n}\tau} & p_{1,n}^-e^{\lambda_{1,n}\tau} & 
			p_{2,n}^+e^{-\lambda_{2,n}\tau} & p_{2,n}^-e^{\lambda_{2,n}\tau} \\ 
		q_{1,n}^+e^{-\lambda_{1,n}\tau} & q_{1,n}^-e^{\lambda_{1,n}\tau} & 
			q_{2,n}^+e^{-\lambda_{2,n}\tau} & q_{2,n}^-e^{\lambda_{2,n}\tau} \\ 
	\end{pmatrix}  
	\begin{pmatrix}
		X_{0,n} \\ X_{1,n} \\ X_{2,n} \\ X_{3,n}
	\end{pmatrix} + 
	\begin{pmatrix}
		Z_{1,n}^- \\ Z_{2,n}^-\\ Z_{1,n}^+\\ Z_{2,n}^+
	\end{pmatrix},
 \label{eq:P3_nlayer_flx}
\end{equation}
where $p_{1,2,n}^\pm = \pi(1\pm 2R_{1,2,n} + \frac{5}{4}Q_{1,2,n})$,
$q_{1,2,n}^\pm = \pi(-\frac{1}{4}+ \frac{5}{4}Q_{1,2,n} \pm 2S_{1,2,n})$ \href{https://github.com/natashabatalha/picaso/blob/9d4cbd672a75c1faf5297c3f1d74074018cd7ef3/picaso/fluxes.py#L3277-L3284}{\faCode}, and
\begin{align}
    Z_{1,n}^\pm(\tau) &= \frac{\pi(1-w_{0,n})}{a_{0,n}}(B_{1,n}\tau + B_{0,n} \pm \frac{2}{a_{1,n}}B_{1,n})),\\
    Z_{2,n}^\pm(\tau) &= -\frac{\pi(1-w_{0,n})}{4a_{0,n}}(B_{1,n}\tau + B_{0,n}).
\end{align}
The boundary conditions for the $P_3$ flux problem are \href{https://github.com/natashabatalha/picaso/blob/9d4cbd672a75c1faf5297c3f1d74074018cd7ef3/picaso/fluxes.py#L3319-L3344}{\faCode}
\begin{alignat}{2}
	F_1^-(0) &= 0, &&\quad\qquad f_1^-(0) = 0, \\
	F_n^-(\partial\tau_n) &= F_{n+1}^-(0), &&\qquad f_n^-(\partial\tau_n) = f_{n+1}^-(0),\\
	F_n^+(\partial\tau_n) &= F_{n+1}^+(0), &&\qquad f_n^+(\partial\tau_n) = f_{n+1}^+(0),
 \end{alignat}
 and
\begin{align}
    F_N^+(\tau_N) =\begin{cases} \pi\left(B(\tau_N) + \frac{2}{3}\frac{\partial B}{\partial\tau}(\tau_N)\right),&\quad \text{non-terrestrial}, \\ 
	\pi B(\tau_N) + A_Sf^-(\tau_N), &\quad \text{terrestrial, hard surface},
    \end{cases}
    \label{eq:bottom_flx_BC_P3_1}
\end{align}
\begin{align}   
f_N^+(\tau_N) =\begin{cases} -\frac{\pi B(\tau_N)}{4},&\quad \text{non-terrestrial}, \\ 
	-\frac{\pi B(\tau_N)}{4} + A_Sf^-(\tau_N), &\quad \text{terrestrial, hard surface},
    \end{cases}
    \label{eq:bottom_flx_BC_P3_2}
\end{align}
for $n={1,2,\cdots,N-1}$ \href{https://github.com/natashabatalha/picaso/blob/9d4cbd672a75c1faf5297c3f1d74074018cd7ef3/picaso/fluxes.py#L2915-L2920}{\faCode}, where $A_S$ is the surface reflectivity.
As with the $P_1$ case, the derivation of the bottom boundary conditions \eqref{eq:bottom_flx_BC_P3_1}--\eqref{eq:bottom_flx_BC_P3_2} are derived in Appendix \ref{app:f_BC} from \cite{mihalas1978stellar}.
As for the $P_1$ case, the spherical harmonics flux problem is formulated in \texttt{PICASO} by representing the system in terms of banded matrices, and solved using the \texttt{solve\_banded} functionality of \texttt{SciPy} \citep{2020SciPy-NMeth}.

\section{Source Function Technique}
\label{sec:source_function}
Following the methodology of \cite{toon1989rapid}, we apply the source function technique to calculate the emergent intensity from the top of the atmosphere.
The radiative transfer equation \eqref{eq:RTE} can be solved to yield the azimuthally integrated intensity at angle $\mu$ at the top of the $n^\text{th}$ layer ($\tau=0$) as
\begin{equation}
	I_n(0,\mu) = I_n(\partial\tau_n,\mu)e^{-\frac{\partial\tau_n}{\mu}} + \frac{1}{\mu}\int_0^{\partial\tau_n} S_{vt}e^{-\frac{\tau}{\mu}}\mathrm{d}\tau,
	\label{eq:azimuth_av_int}
\end{equation}
for
\begin{equation}
	S_{vt} = \frac{w_{0,n}}{2}\int_{-1}^{1} I_t(\tau, \mu')P(\mu,\mu')\mathrm{d}\mu' + S_n(\tau) ,
	\label{eq:Svt}
\end{equation}
where 
\begin{equation}
	S_n(\tau) = 2\pi(1-w_{0,n})(B_{0,n}+B_{1,n}\tau).
\end{equation}
\cite{toon1989rapid} showed that infrared intensities can be estimated with sufficient accuracy by using the two-stream approximation to define the source function in the equation of radiative transfer, therefore, we use $I_t$, the solution to the $P_1$/$P_3$ problem outlined in Section \ref{sec:SH}, in place of the true intensity in the source term \eqref{eq:Svt}.
Therefore we can rewrite \eqref{eq:Svt} as
\begin{equation}
	S_{vt} = w_{0,n}\summ \chi_l I_l(\tau)P_l(\mu) + S_n(\tau). 
	\label{eq:Svt_1}
\end{equation}
Let us consider the integral term in \eqref{eq:azimuth_av_int}. Using \eqref{eq:Svt_1}, this can be written as
\begin{equation}
	\int_0^{\partial\tau_n} S_{vt}e^{-\frac{\tau}{\mu}}\mathrm{d}\tau = 
		w_{0,n}\summ \chi_l P_l(\mu) \int_0^{\partial\tau_n}I_l(\tau) e^{-\frac{\tau}{\mu}}\mathrm{d}\tau 
		+ \int_0^{\partial\tau_n}  S_n(\tau) e^{-\frac{\tau}{\mu}}\mathrm{d}\tau.
		\label{eq:Svt_int}
\end{equation}
We can  calculate the second term on the right-hand side of  \eqref{eq:Svt_int} to be \href{https://github.com/natashabatalha/picaso/blob/9d4cbd672a75c1faf5297c3f1d74074018cd7ef3/picaso/fluxes.py#L2812-L2814}{\faCode}
\begin{equation}
	\int_0^{\partial\tau_n}  S_n(\tau) e^{-\frac{\tau}{\mu}}\mathrm{d}\tau=
    2\pi\mu(1-w_{0,n}) \left[B_{0,n}\left(1 - e^{-\frac{\partial\tau}{\mu}}\right)
    + B_{1,n}\left(\mu - \left(\partial\tau + \mu\right)e^{-\frac{\partial\tau}{\mu}}\right)\right]
    \label{eq:integrated_source}
\end{equation}
Next, let us write $\Aint = \int_0^{\partial\tau_n}I_l(\tau) e^{-\frac{\tau}{\mu}}\mathrm{d}\tau $. 
This is calculated identically for reflected light, and is outlined in detail in \cite{rooney2023spherical}, where the solution is given by
\begin{equation}
	\Aint = A_nX_n + N_n 
	\label{eq:Aint}
\end{equation}
where matrix $A_n$ is defined in \cite{rooney2023spherical} \href{https://github.com/natashabatalha/picaso/blob/9d4cbd672a75c1faf5297c3f1d74074018cd7ef3/picaso/fluxes.py#L2999-L3002}{\faCode} and $X_n$ are the coefficients we solve for in the $P_1$ and $P_3$ flux problems \eqref{eq:P1_nlayer_flx}--\eqref{eq:bottom_flx_BC_P1} and \eqref{eq:P3_nlayer_flx}--\eqref{eq:bottom_flx_BC_P3_2} respectively.
Vector $N_n$ differs from that for reflected light. 

For infrared sources, $N_n$ is defined as
\begin{align}
	N_{0,n} &= \frac{2\pi\mu \left(1-w_{0,n}\right)}{a_{0,n}}\left[ B_{0,n}\left(1-e^{-\frac{\partial\tau}{\mu}}\right) + B_{1,n}\left(\mu - \left(\partial\tau+\mu\right)e^{-\frac{\partial\tau}{\mu}}\right)\right], \\
     N_{1,n} &= \frac{2\pi\mu (1-w_{0,n})}{a_{0,n}} \frac{B_{1,n}}{a_{1,n}}\left(1 - e^{-\frac{\partial\tau}{\mu}}\right),\\
     N_{2,n}&=N_{3,n}=0.
\end{align}
Substituting $\Aint$ \eqref{eq:Aint} and the integrated source term \eqref{eq:integrated_source} back into \eqref{eq:Svt_int}, we can use \eqref{eq:azimuth_av_int} to calculate the azimuthally integrated intensity emerging from the top of the $n^\text{th}$ layer. 
By beginning at the bottom of the atmosphere ($n=N$) and working our way up layer-by-layer, we can derive the azimuthally integrated intensity at the top of the atmosphere.
This intensity is used to calculate the infrared flux to predict the observed atmospheric spectra.

\section{Analysis}
\label{sec:analysis}

To quantitatively analyze the performance of the spherical harmonics method for infrared radiative transfer, we compare our results with Toon89 and \texttt{CDISORT}, a version of the discrete ordinate solver, \texttt{DISORT}, written in C rather than FORTRAN \citep{stamnes1988numerically,stamnes2000disort, mayer2005libradtran, buras2011new}.
\texttt{CDISORT} is a versatile, well-tested and widely used radiative transfer software, with advanced numerical capabilities. 
\texttt{CDISORT} has the capacity to model $L$-stream discrete ordinates approximations, where $L$ is arbitrary and considerably greater than 4 (we will study 16 and 32 stream calculations in this work).

One important note to consider before delving into comparisons of Toon89 and \texttt{CDISORT} is the different scattering phase functions. The Toon89 methodology utilizes the hemispheric mean phase function for infrared scattering \citep{toon1989rapid}.
The hemispheric mean approach is derived by assuming that the phase function takes the value of $1+g_0$ in the forward scattering hemisphere, and $1-g_0$ in the backward scattering hemisphere, where $g_0$ denotes the asymmetry parameter.
\cite{toon1989rapid} chose this technique because for infrared wavelengths, it assumes the correct relationship between flux and intensity and produces the proper emissivity in the limiting case of dominant absorption ($w_0=0$) for a semi-infinite atmosphere.

On the other hand, \texttt{CDISORT}, SH2 and SH4 all utilize the Henyey-Greenstein phase function \citep{henyey1941diffuse}:
\begin{equation}
	P_\text{HG}(\cos\Theta) = \frac{1-g_0^2}{(1+g_0^2-2g_0\cos\Theta)^{3/2}},
	\label{eq:HG}
\end{equation}
where the scattering angle $\Theta$ is defined as
\begin{align}
	\cos\Theta &= \mu\mu' - \sqrt{1-\mu^2}\sqrt{1-\mu'^2}\cos(\phi-\phi').
	\label{eq:costheta}
\end{align}
for incoming and outgoing radiation angular directions $(\mu,\phi)$ and $(\mu',\phi')$ respectively. 

We emphasize this difference in computational methods to foreshadow differences that arise between the methodologies. In what follows, we first compare two benchmark spectra in \S\ref{sec:sample_atmos}. Then, we isolate the dependence of each method's accuracy on scattering parameters (single scattering, asymmetry) in \S\ref{sec:heatmaps}. Lastly, we baseline the timing of these methodologies in \S\ref{sec:timing_accuracy} to investigate the trade-off between computational time and accuracy. 

\subsection{Comparison of benchmark spectra}
\label{sec:sample_atmos}
\begin{figure}[t!]
    \centering
    \gridline{\fig{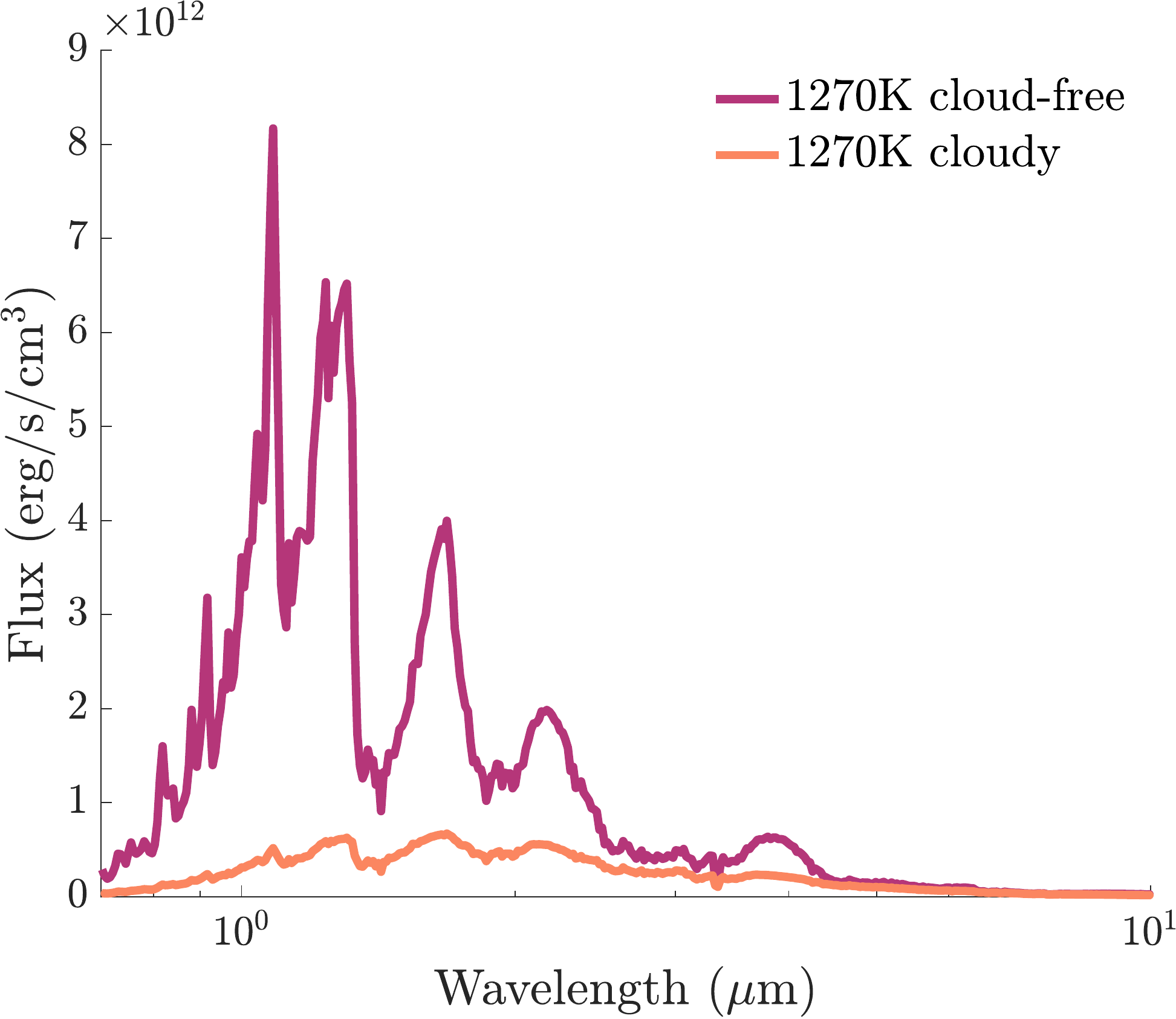}{0.48\textwidth}{(a) $T_\text{eff}=1270K$.}
    \fig{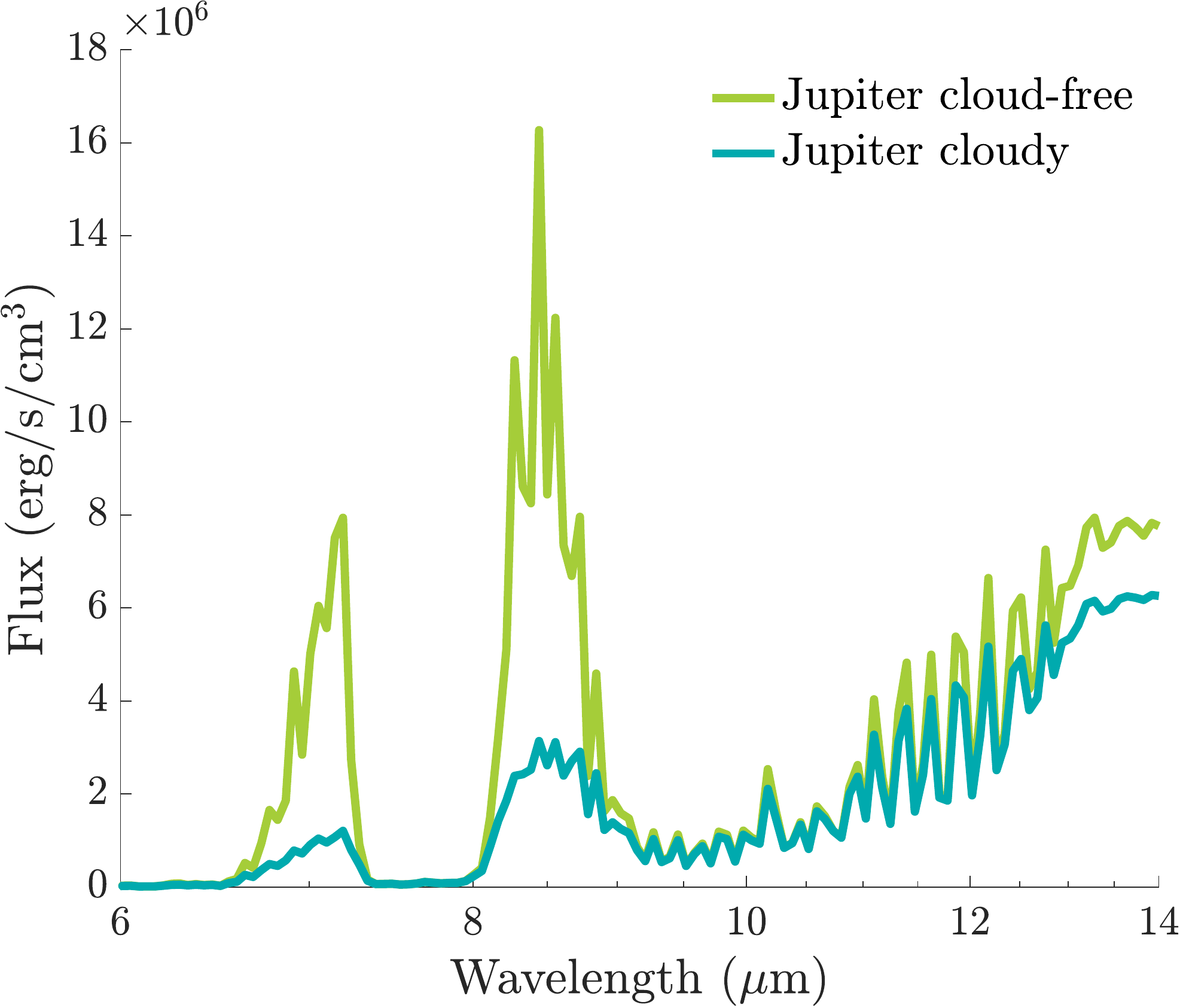}{0.48\textwidth}{(b) Jupiter-like.}
    }
    \caption{Infrared spectra with and without clouds predicted by Toon89 in \texttt{PICASO} for two different atmospheres: (a) brown dwarf with effective temperature $T=$1200 K, gravity $g=$200 m/s$^2$, solar metallicity, solar C/O and forsterite, iron, corundum clouds; (b) planet similar to Jupiter with $g=$25m/s$^2$, semi major axis=5 AU, orbiting a Sun-like star, with H$_2$O and NH$_3$ clouds.}
	\label{fig:picaso_spectra}
\end{figure}

We consider two different benchmark atmospheres on which to conduct our analysis: (i) a brown dwarf with effective temperature $T_\text{eff}=1200$ K, gravity $g=200$ m/s$^2$, solar metallicity, solar C/O and forsterite, iron, corundum clouds, and (ii) planet similar to Jupiter with $g=$25m/s$^2$, semi major axis=5 AU, orbiting a Sun-like star, with H$_2$O and NH$_3$ clouds.
The cloudy and non-cloudy infrared spectra, as predicted by the \cite{toon1989rapid} implementation in \texttt{PICASO} \citep{natasha_batalha_2022_6419943}, is denoted Toon89.

As shown in Figure \ref{fig:picaso_spectra} the cases chosen both have spectra that are largely affected by the presence of clouds. In both cases, the clouds act to prevent photon contribution from the deepest, hottest, layers. As a result, the pressure range probed by the cloudy models is limited to the upper, cooler layers, creating spectral features that appear muted, relative to the cloud-free counterpart. We choose these cases in order to test the accuracy of these methodologies in the scattering-dominated limit for typical cloudy objects. We note that the different methodologies agree in the case of no scattering, indicating that any spectral differences in the cloudy cases are a consequence of the approximations implemented to deal with scattering.

\begin{figure*}[t!]
\centering
	\gridline{\fig{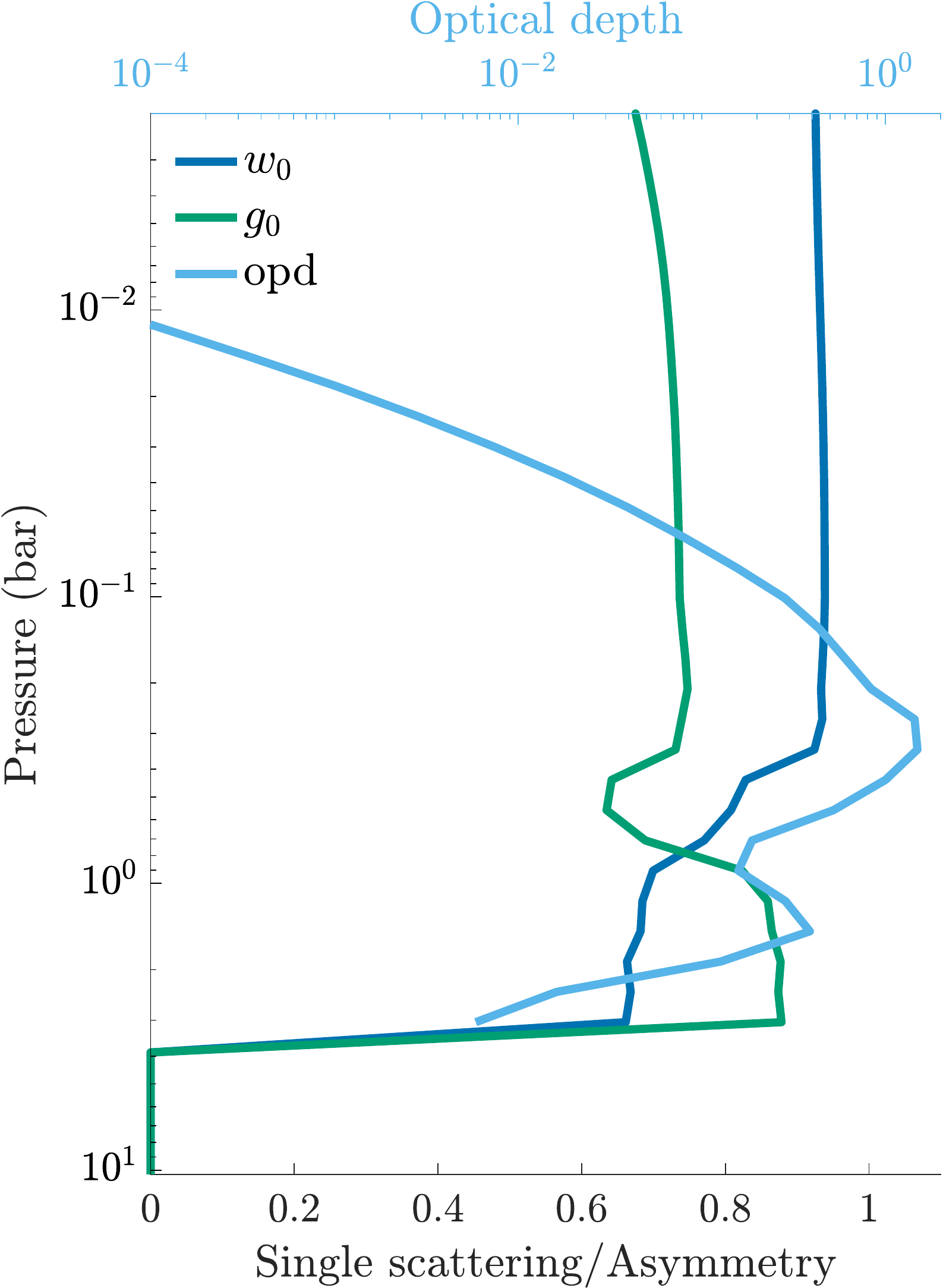}{.3\textwidth}{(a)}
	\fig{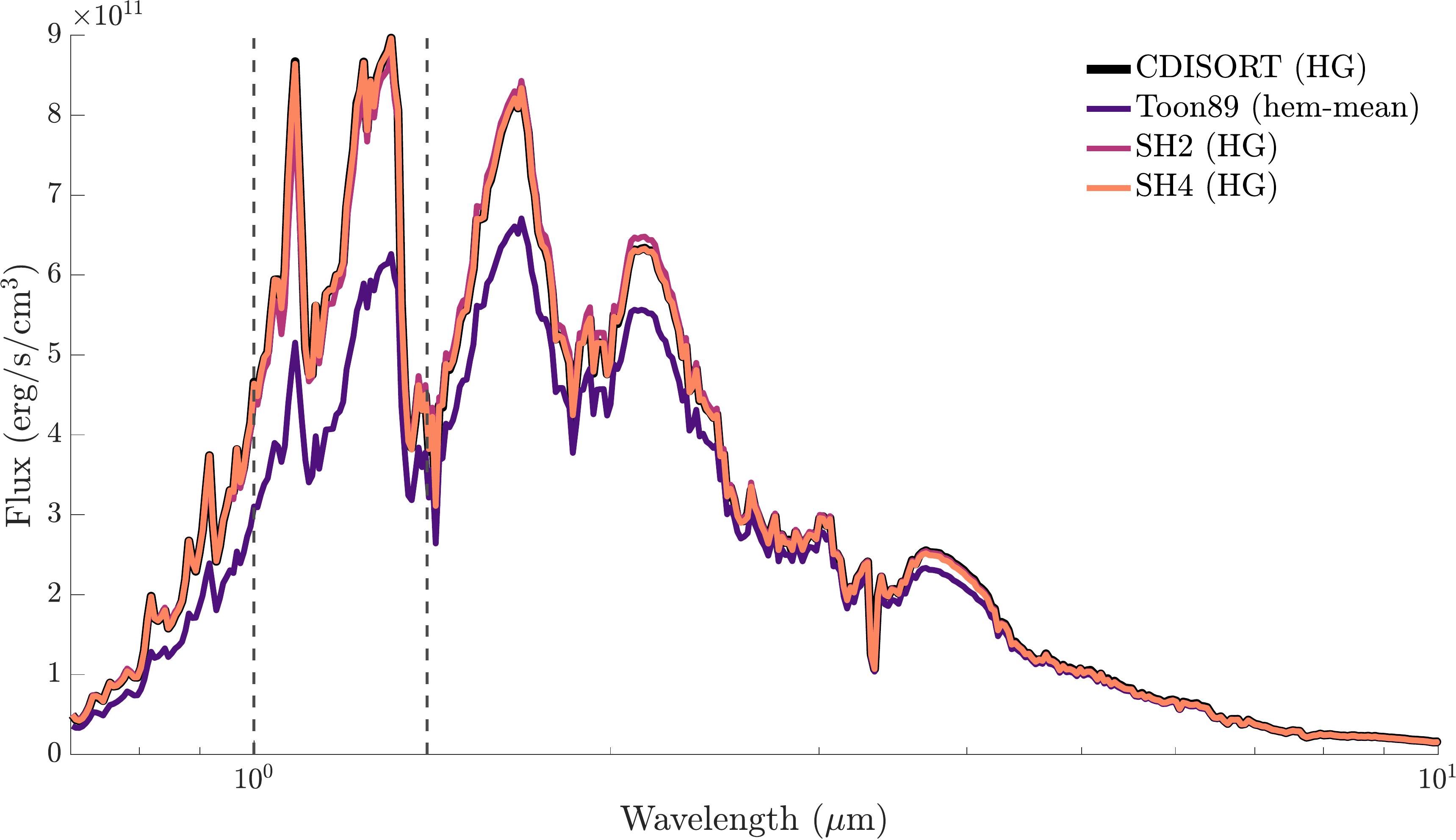}{.67\textwidth}{(b) }}
	\caption{Comparison between the infrared spectra predicted by 16-stream \texttt{CDISORT}, \texttt{PICASO}, 2-term spherical harmonics (SH2) and 4-term spherical harmonics (SH4) for a cloudy brown dwarf with effective temperature $T=$1200 K, gravity $g=$200 m/s$^2$, solar metallicity, solar C/O and forsterite, iron, corundum clouds. We use HG and hem-mean to indicate that the model phase function is Henyey-Greenstein or hemispheric mean respectively. The scattering properties plotted in subfigure (a) correspond to the average values within the 1-1.4$\mu$m wavelength region, as marked by the grey dashed lines on the spectra plot (b). \href{https://github.com/natashabatalha/picaso/blob/9d4cbd672a75c1faf5297c3f1d74074018cd7ef3/docs/notebooks/10c_AnalyzingApproximationsThermal.ipynb}{\faCode}}
	\label{fig:cloudy_1270K_comparison}
\end{figure*}

In Figure \ref{fig:cloudy_1270K_comparison}(b) we plot the infrared spectra for cloudy atmosphere (i), predicted by 16-stream \texttt{CDISORT}, Toon89, two-term spherical harmonics (SH2) and four-term spherical harmonics (SH4).
Note that we indicate whether the models utilise the Henyey-Greenstein (HG) or hemispheric mean (hem-mean) phase function in the figure legend.
Figure \ref{fig:cloudy_1270K_comparison}(b) depicts the single scattering, asymmetry and optical depth profiles with pressure, averaged in the wavelength range 1--1.4$\mu$m, as indicated by the grey dashed lines on the spectra plot.
We choose this wavelength window to average the scattering parameters as it is a region with significant differences in the spectra produced by Toon89 and the other models.
The cloud profile is shaped by two cloud layers: one smaller cloud layer below 1~bar overlayed by a larger optical depth cloud layer above 1~bar. In the high optical depth region ($\tau>$0.5), the associated optical properties range between 0.6--0.75 for the asymmetry and 0.8--0.94 for the single scattering. These values are typical for these condensate species and present a less forward scattering example, as compared with the ``Jupiter-like'' example. 

\begin{figure}[b!]
    \centering
    \epsscale{.9}
    \plotone{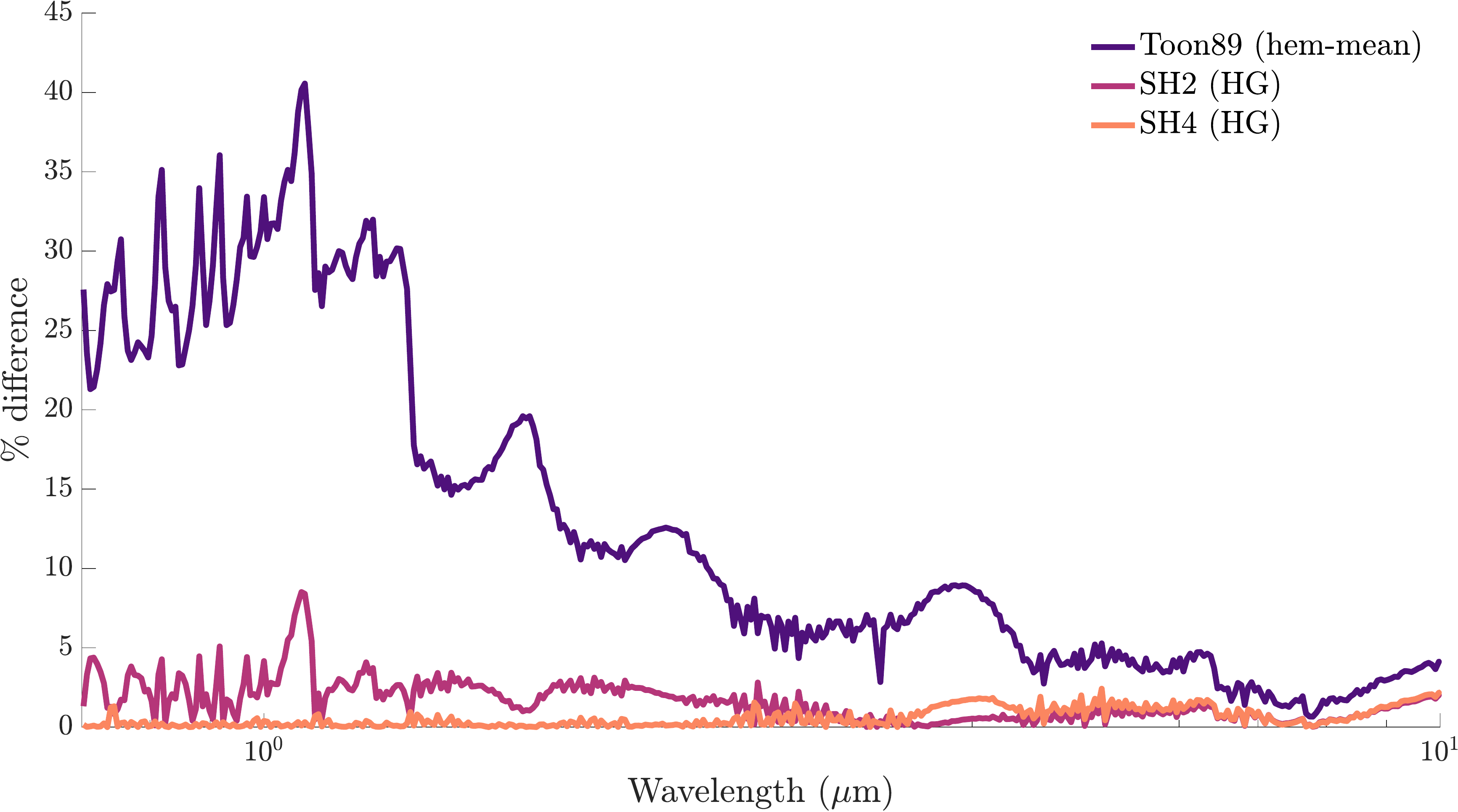}
    \caption{Percentage differences between the spectra produced by \texttt{CDISORT} and Toon89, SH2 and SH4 for the $T_\text{eff}=1270K$ profile, where HG and hem-mean indicates that the model phase function is Henyey-Greenstein or hemispheric mean respectively.}
	\label{fig:cloudy_1270K_diff}
\end{figure}

There is an immediately noticeable difference between Toon89, \texttt{CDISORT} and SH2/4. Given the close agreement of SH2 with 16-stream \texttt{CDISORT} compared to Toon89, which is also a two-stream technique, we can isolate that it is the choice of phase function that leads to this difference.
\cite{rooney2023spherical} conducted an investigation into the accuracy gain when applying four-term spherical harmonics to predict the scattering of reflected light in atmospheres, and compared geometric albedo produced by SH2, SH4 and Toon89 with the doubling method, calculated by \cite{liou1973numerical}.
The authors concluded that SH2 and Toon89 performed comparably, and that the choice between spherical harmonics or discrete-ordinates had little impact on the solution accuracy when compared to the doubling method.
The only difference between the SH2 method applied in this work and that applied in \cite{rooney2023spherical} is the thermal source and boundary conditions, which are identically applied to Toon89 in \texttt{PICASO}. 
However, the \texttt{PICASO} implementation of Toon89 for reflected light leverages a post-processed Henyey-Greenstein phase function for direct scattering, opposed to the hemispheric mean approach used in infrared \citep{natasha_batalha_2022_6419943}.

For a clearer understanding of how SH4 compares to SH2, we also plot the percentage difference between 16-stream \texttt{CDISORT} and Toon89, SH2 and SH4 in Figure \ref{fig:cloudy_1270K_diff}.
We notice that the largest deviance of SH2 from 16-stream \texttt{CDISORT} is around 8.5\%, whereas SH4 is always within 2.5\%.

\begin{figure}[t!]
\centering
	\gridline{\fig{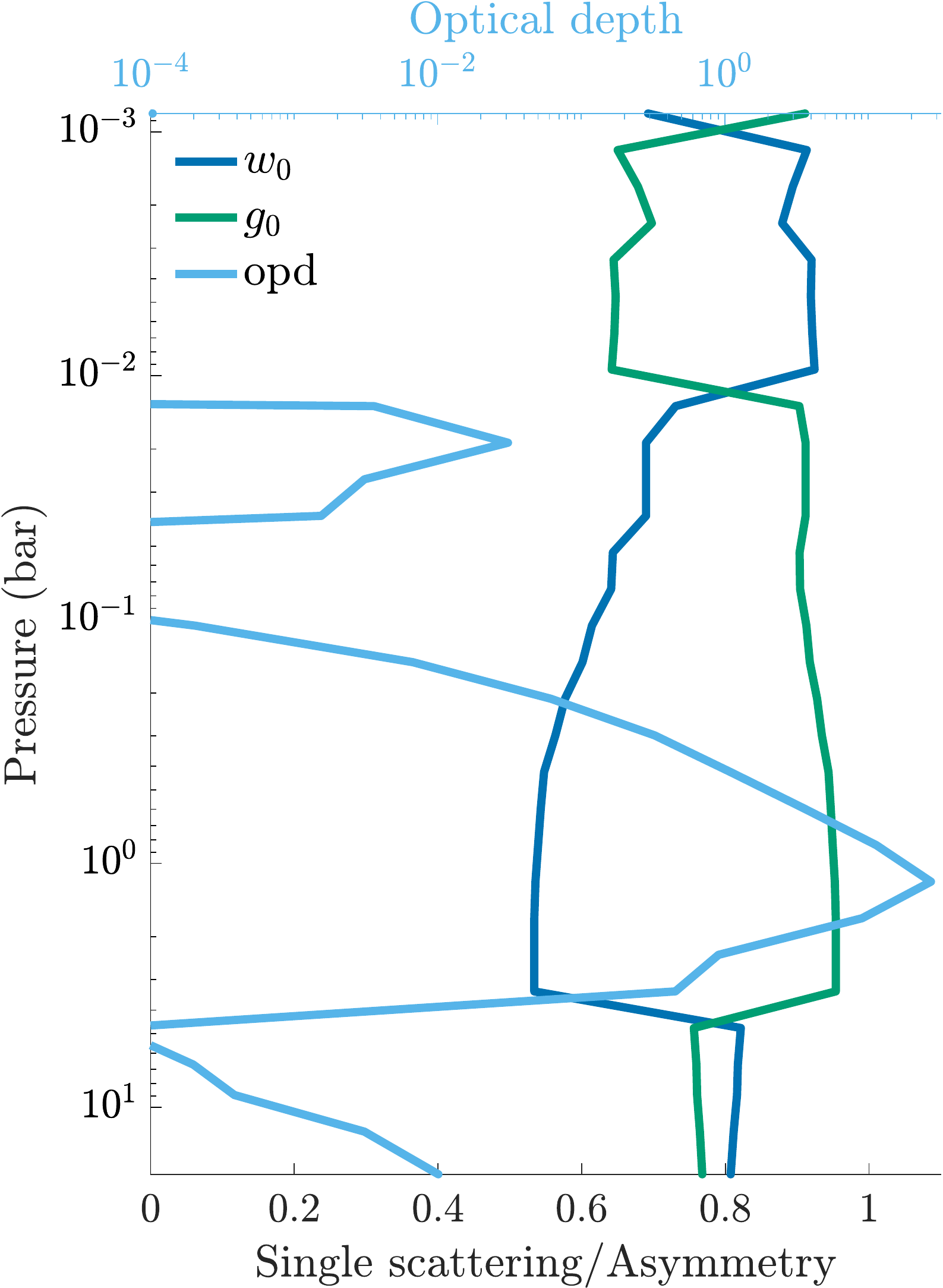}{.3\textwidth}{(a)}
	\fig{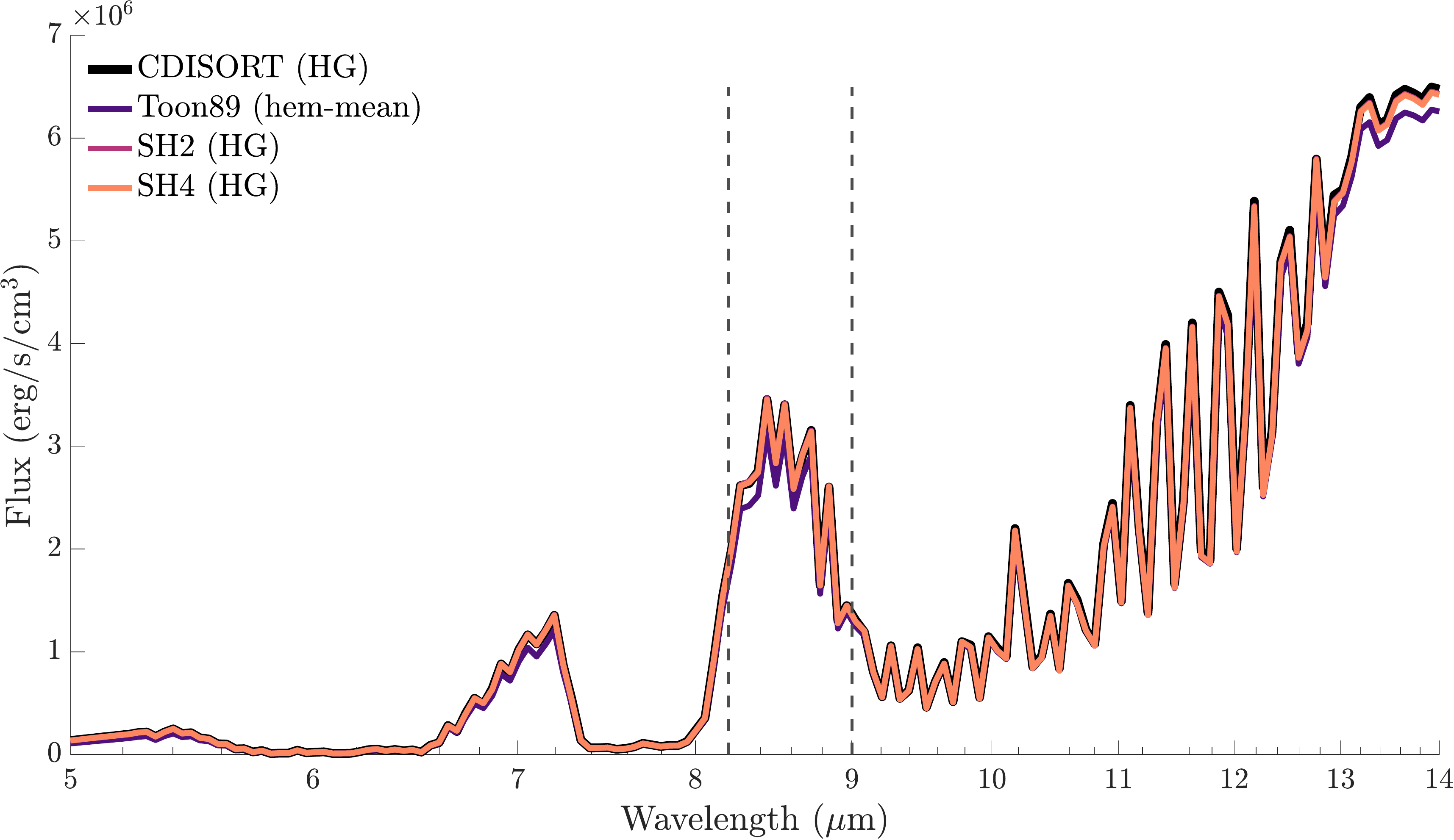}{.67\textwidth}{(b) }}
	\caption{Comparison between the infrared spectra predicted by 16-stream \texttt{CDISORT}, \texttt{PICASO}, 2-term spherical harmonics (SH2) and 4-term spherical harmonics (SH4) for a Jupiter-like profile. The scattering properties plotted in subfigure (a) correspond to the average values within the 8.2-9$\mu$m wavelength region, as marked by the grey dashed lines on the spectra plot (b). \href{https://github.com/natashabatalha/picaso/blob/9d4cbd672a75c1faf5297c3f1d74074018cd7ef3/docs/notebooks/10c_AnalyzingApproximationsThermal.ipynb}{\faCode}}
	\label{fig:cloudy_jupiter_comparison}
\end{figure}

\begin{figure}[b!]
    \centering
    \epsscale{.9}
    \plotone{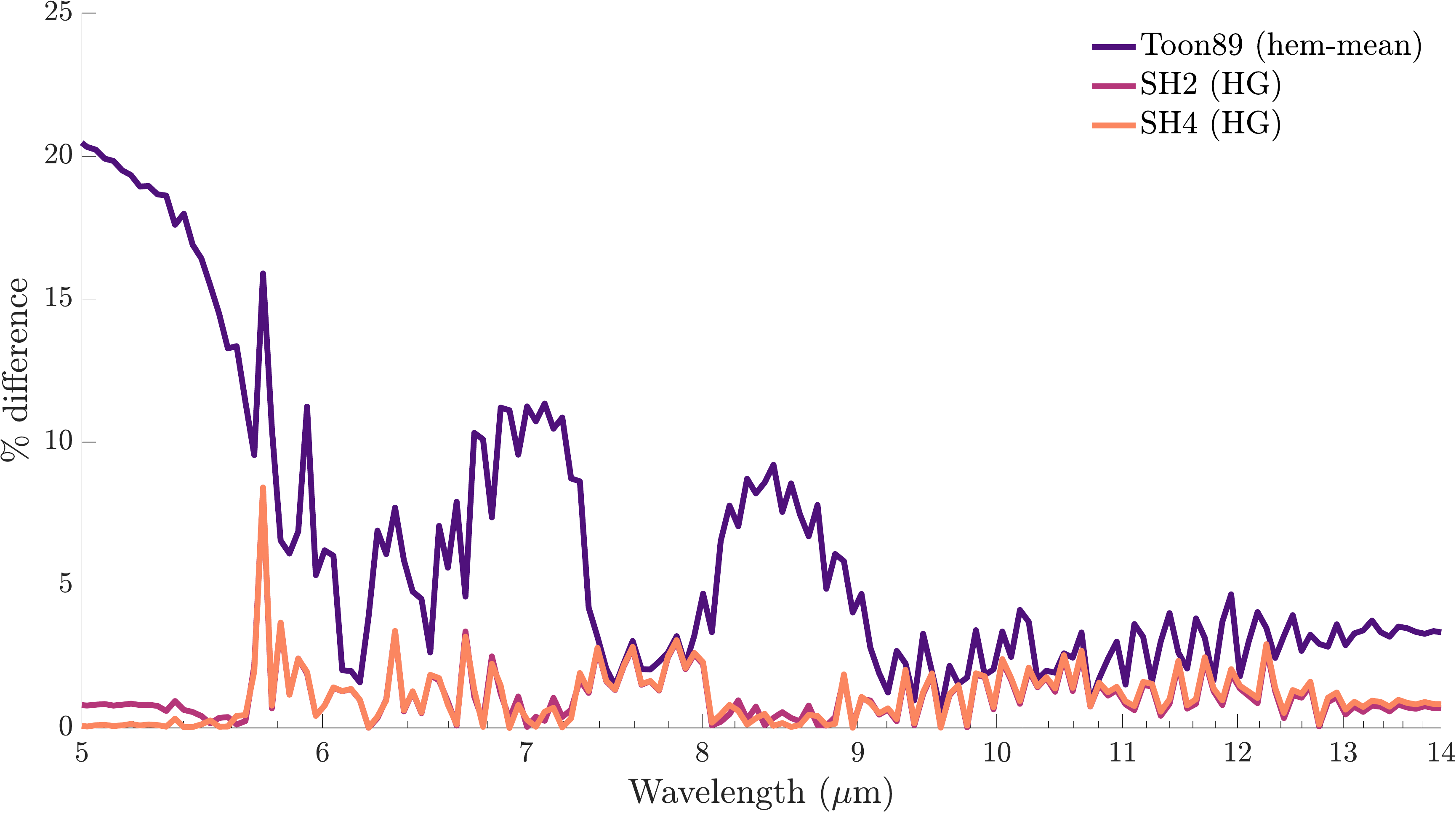}
    \caption{Differences in spectra for cloudy Jupiter.}
	\label{fig:cloudy_jupiter_diffs}
\end{figure}

We conduct the same analysis for the Jupiter-like profile in Figure \ref{fig:cloudy_jupiter_comparison}, where the infrared spectra predicted by 16-stream \texttt{CDISORT}, Toon89, SH2 and SH4 are plotted on the right, alongside the single scattering, asymmetry and optical depth profiles with pressure, averaged in the wavelength range 8.2--9$\mu$m.
We notice immediately that the four models are in close agreement throughout the entire wavelength range, with the greatest differences evident between 8--9$\mu$m and for wavelengths greater than 13$\mu$m.
We attribute this to the different cloud condensate optical properties of the Jupiter-like profile. The cloud profile in this case is also shaped by two cloud layers: one larger cloud deck around  1~bar and another smaller optical depth cloud layer around 0.02~bar. In the high optical depth region, the associated optical properties range between 0.8--0.94 for the asymmetry and 0.53--0.8 for the single scattering. Since this atmosphere exhibits greater forward scattering but with lower values for the single scattering albedo, this suggests that the accuracy of SH4, SH2, and Toon89 methods are highly dependent on what the strength of scattering and asymmetry of the cloud is.

We again plot the percentage difference between 16-stream \texttt{CDISORT} and Toon89, SH2 and SH4 in Figure \ref{fig:cloudy_jupiter_diffs}, and observe a maximum deviance around 11\% for Toon89 (ignoring wavelengths less than 6.6$\mu$m where spectra values themselves are very small).
We also notice that SH2 and SH4 exhibit effectively identical agreement with 16-stream \texttt{CDISORT}, with percentage differences staying below 3.5\%.

\subsection{Dependence of accuracy on scattering parameters}
\label{sec:heatmaps}

\begin{figure}[b!]
    \centering
    \gridline{\fig{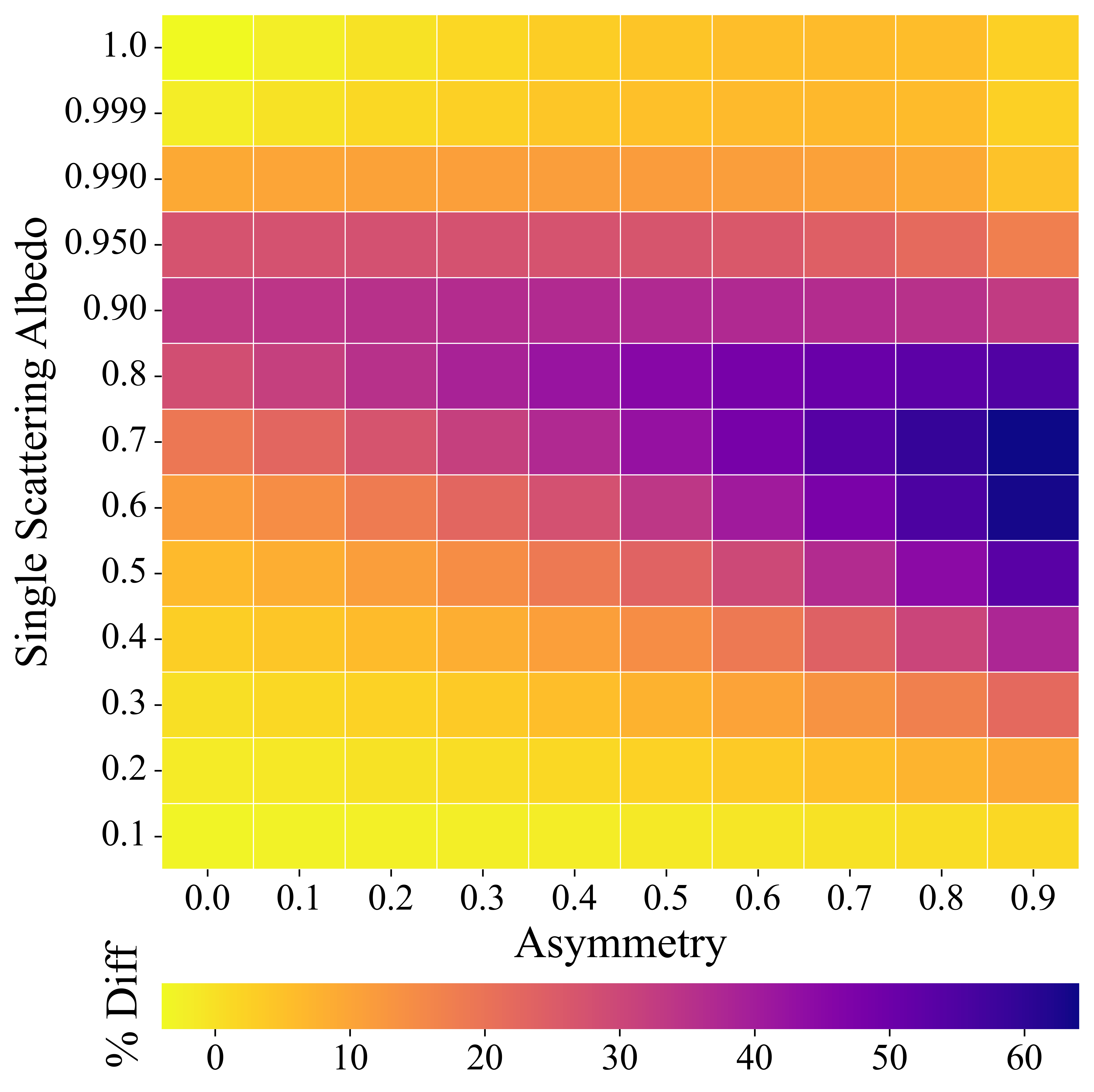}{0.33\textwidth}{(a) Toon89}\hspace{-1cm}
    \fig{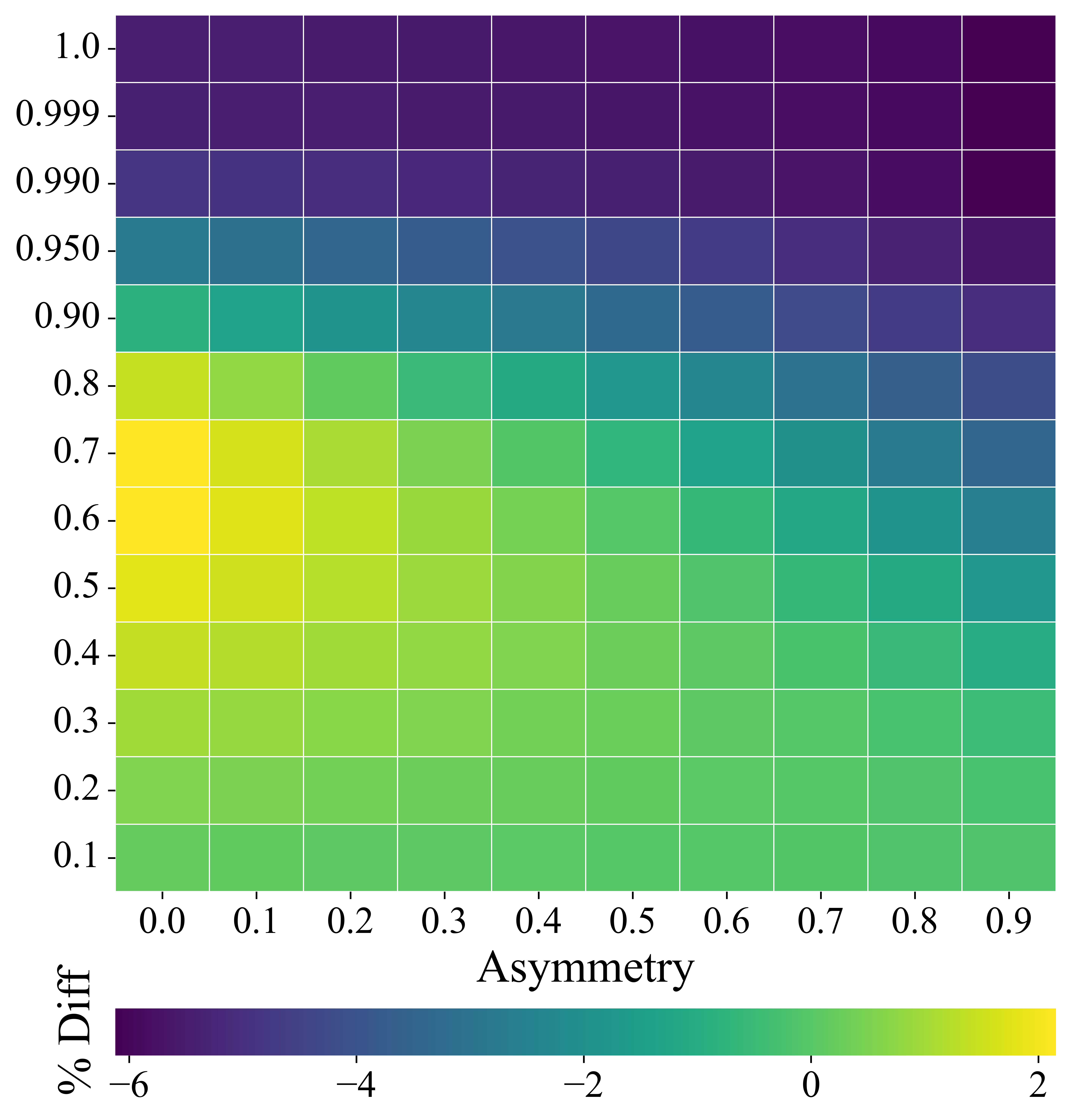}{0.316\textwidth}{(b) SH4}\hspace{-1cm}
    \fig{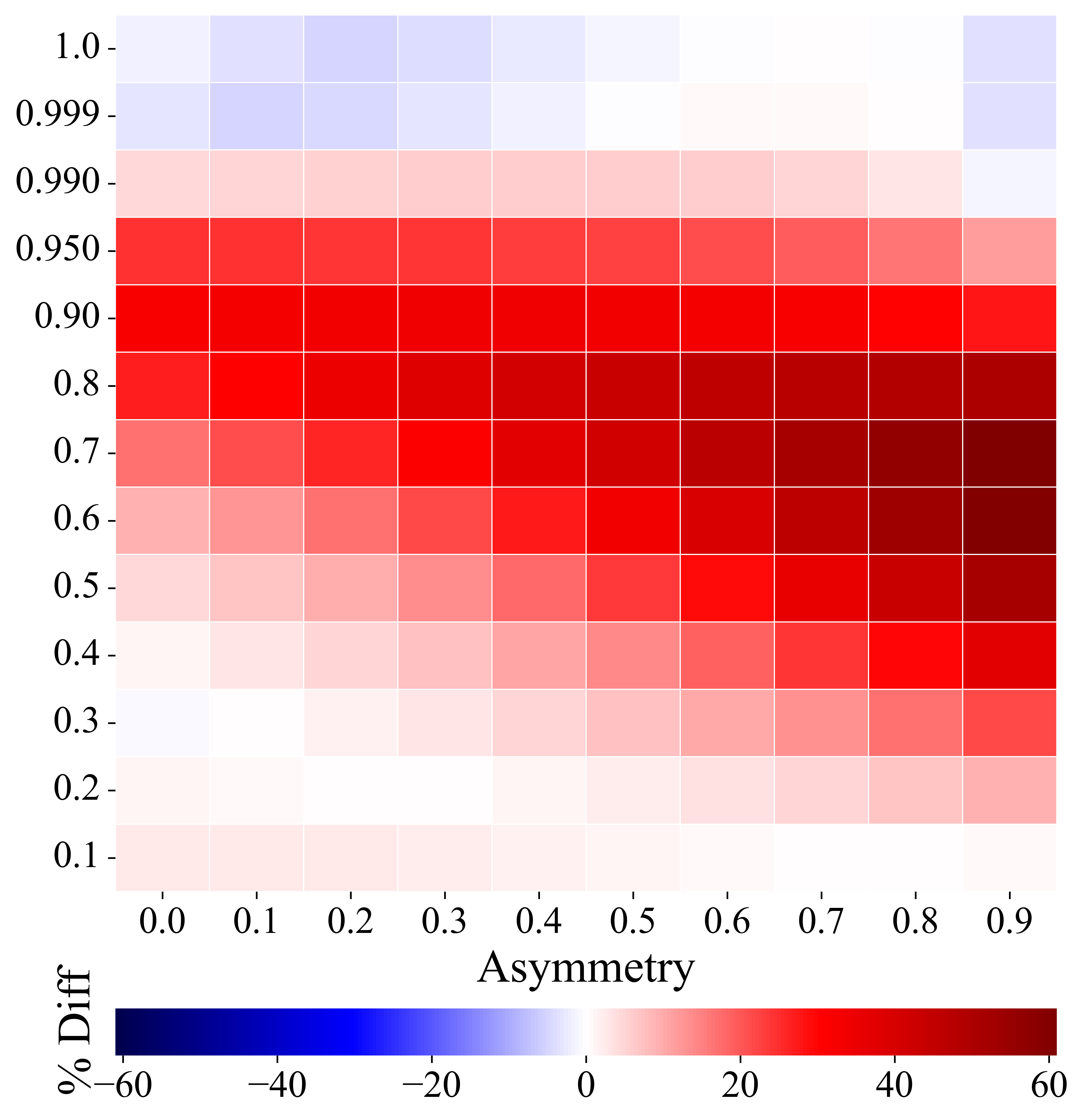}{0.32\textwidth}{(c) Difference between figs (a) and (b)}
    }
    \caption{Heatmaps depicting the percentage difference in average flux produced by (a) Toon89 and (b) SH4 with 32-stream \texttt{CDISORT}. Figure (c) is produced by subtracting the absolute percentage differences of SH4 from that of Toon89 to elucidate for which parameters one method agree with \texttt{CDISORT} better than the other. Dark-red represents cases where SH4 outperforms Toon89, as compared to \texttt{CDISORT}. White (toward zero) represents cases where Toon89 and SH4 perform comparably. \href{https://github.com/natashabatalha/picaso/blob/9d4cbd672a75c1faf5297c3f1d74074018cd7ef3/docs/notebooks/10c_AnalyzingApproximationsThermal.ipynb}{\faCode} }
	\label{fig:heatmap}
\end{figure}

Since it is clear that there is a large accuracy dependence on single scattering and asymmetry, we compare the spectra produced by Toon89 and SH4 with 32-stream \texttt{CDISORT} for a range of single-scattering albedos and asymmetry parameters in a test atmosphere.
We define the test atmosphere with the same pressure-temperature and optical depth profile as the $T_\text{eff}=1270$K case studied in Section \ref{sec:sample_atmos}, however we force the cloud single-scattering albedo $w_0$ and asymmetry parameter $g_0$ to take constant values for clarity.
We sweep over a grid of parameter values and calculate the resulting spectra for each of our three models. 
We consider $w_0$ in the range 0.1--1.0, and $g_0$ in the range 0.0--0.9. 
Note that we consider a finer grid for high ($w_0>$0.9) single-scattering albedo.
Taking an average of the spectra values over the wavelength range 1--10$\mu$m for each pair of $w_0$ and $g_0$ values, we calculate the percentage difference between 32-stream \text{CDISORT} and each of Toon89 and SH4. 
We plot the results as heatmaps in Figure 
\ref{fig:heatmap}, where Figure \ref{fig:heatmap}(a) depicts the percentage difference in infrared flux between Toon89 and 32-stream \texttt{CDISORT}, and Figure \ref{fig:heatmap}(b) displays that of SH4 and \texttt{CDISORT}.
To further elucidate for which parameters SH4 and Toon89 better agree with \texttt{CDISORT}, we subtract the absolute percentage difference of SH4 with \texttt{CDISORT} from that of Toon89, and plot the result in Figure \ref{fig:heatmap}(c).
In the red-colored regions, SH4 out-performs Toon89 when compared to \texttt{CDISORT}. 
The white regions represent the cases where SH4 and Toon89 have comparable agreement with \texttt{CDISORT}. The blue-colored regions represent when Toon89 is in closest agreement with \texttt{CDISORT}.

We see from Figure \ref{fig:heatmap} that Toon89 exhibits a maximum percentage difference of around 60\% with 32-stream \texttt{CDISORT} for high asymmetry ($g_0>0.8$) and moderate single-scattering albedo ($0.5<w_0<0.8$).
The lowest errors occur for extreme values of single scattering, namely $w_0=0.1$ and $w_0>0.99$. 
We note the excellent agreement for $w_0=0$, which validates the justification of using the hemispheric mean phase function by \cite{toon1989rapid} to ensure correct emissivity in the $w_0=0$ limit.

Superior agreement is achieved by SH4 when compared with 32-stream \texttt{CDISORT}, with the maximum error of -6\% occurring for single-scattering albedo exceeding 0.95.
By comparing the two heatmaps in this region, we see that Toon89, using the hemispheric mean approximation, agrees more closely with 32-stream \texttt{CDISORT} than SH4, even though both SH4 and \texttt{CDISORT} use the Henyey-Greenstein phase function.
This implies that for high single-scattering ($w_0=1$), the hemispheric-mean approximation marginally outperforms low-order spherical harmonic expansions for the Henyey-Greenstein phase function.

\subsection{Timing-accuracy trade-off}
\label{sec:timing_accuracy}
\begin{figure}[b!]
    \centering
    \gridline{\fig{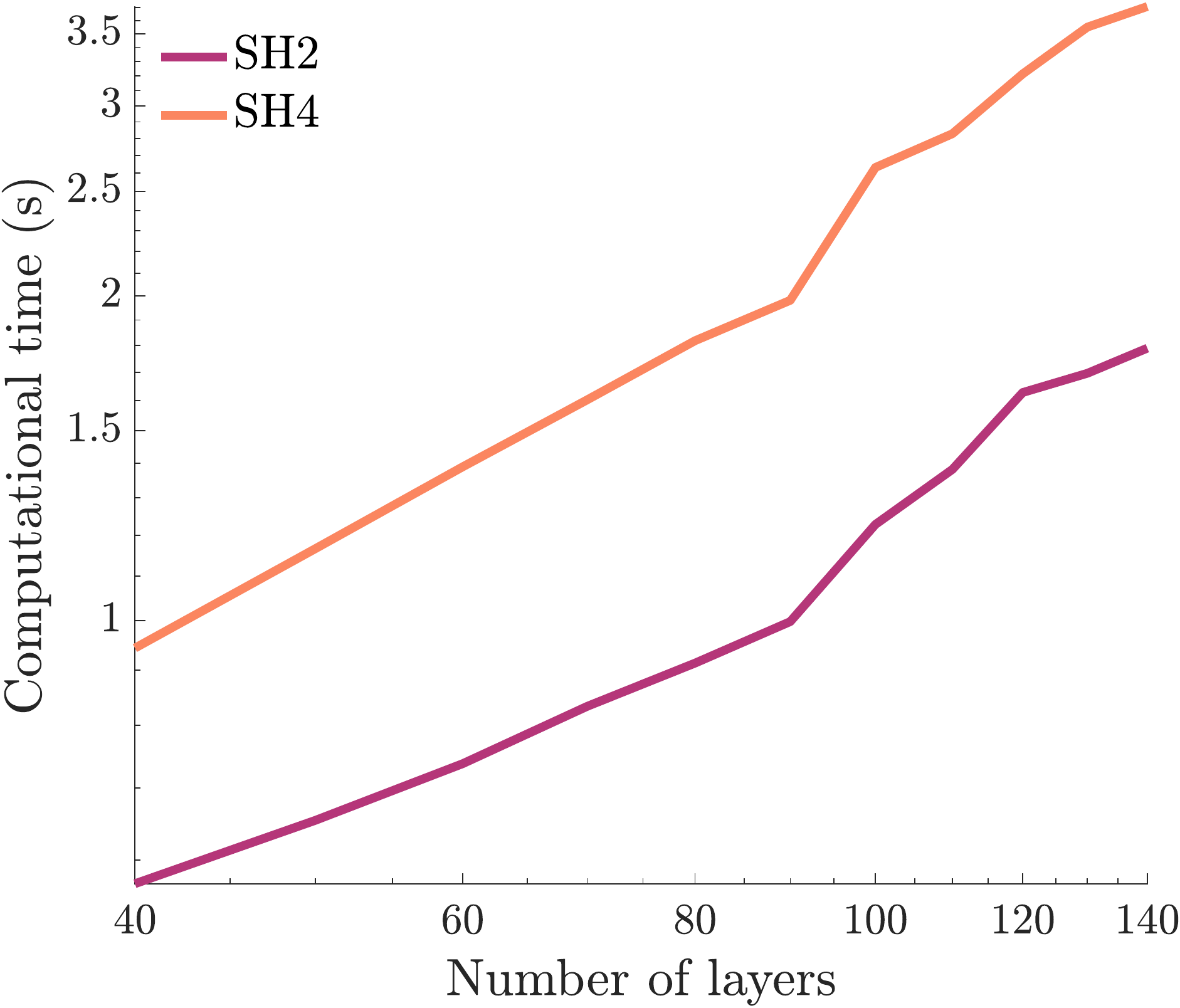}{0.45\textwidth}{(a) Computational time.}
    \fig{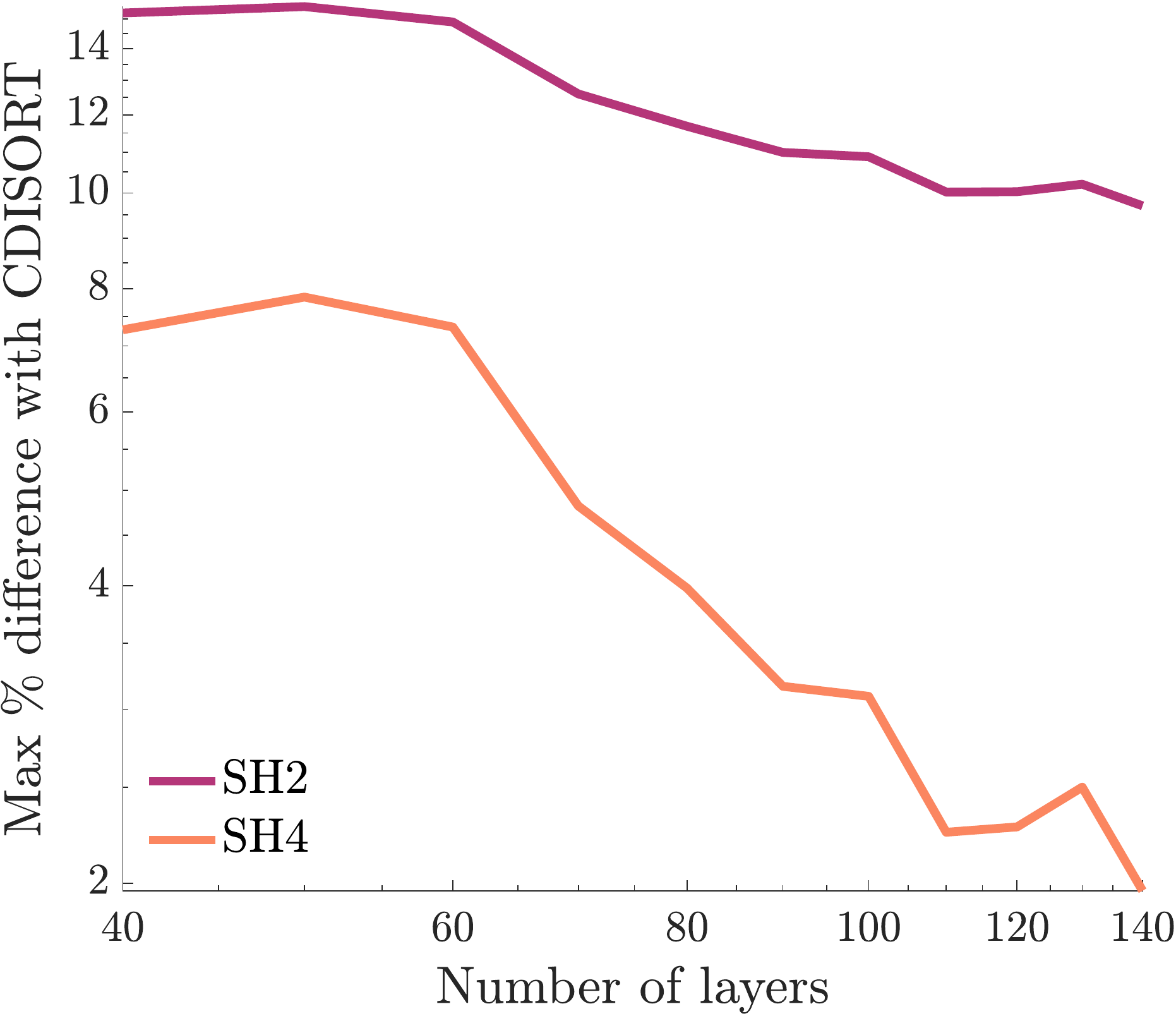}{0.45\textwidth}{(b) Maximum percentage difference with 16-stream \texttt{CDISORT}.}
    }
    \caption{Analyzing how the (a) computational time and (b) maximum percentage difference of SH2 and SH4 with \texttt{CDISORT} changes as the number of layers is increased from 40 to 140. We use a 16-stream, 140-layer \texttt{CDISORT} model to benchmark the spherical harmonics against. We see an evident increase in computational expense with number of layers, where SH4 is twice as slow as SH2 for the 140-layer case, however, the maximum percentage difference with the benchmark decreases significantly with layers for SH4, where SH4 is within 2\% of \texttt{CDISORT} versus 9.7\% for SH2.}
	\label{fig:timing_accuracy}
\end{figure}

Despite the improvements of moving to SH4, we still must consider the timing-accuracy trade-off. 
To elucidate this, we analyze the computational expense of SH2 and SH4 as the number of layers is increased from 40 to 140, alongside the maximum percentage difference of their thermal spectra with that of a 16-stream, 140-layer \texttt{CDISORT} model.
We run this analysis on the $T_\text{eff}=1270$K atmosphere studied above, over a wavelength range of 0.7--2$\mu$m and plot our results in Figure \ref{fig:timing_accuracy}.

As the focus of the present study is to assess the trade-off between computational expense and model accuracy, we compare only SH2 and SH4 to \texttt{CDISORT} to illustrate the increase in cost and improvement in agreement with higher-fidelity models when moving from two to four terms.
The modelling approach of SH2 and SH4 is identical bar the number of terms, whereas the Toon89 methodology differs both in model choice (discrete-ordinates versus spherical harmonics) and phase function (hemispheric mean versus Henyey-Greenstein).
In an attempt to attribute any differences in computational expense and agreement with \texttt{CDISORT} to only the number of layers chosen for the model, we compare SH2 with SH4.

From Figure \ref{fig:timing_accuracy}(a) we see an increase in computational expense, $t$, for both SH2 and SH4 when the number of layers, $N$, is increased from 40 to 140 scaling approximately as $t=\mathcal{O}(N)$.
Overall, SH4 is approximately twice as expensive as SH2.
However, in Figure \ref{fig:timing_accuracy}(b) SH4 has a significant increase in model agreement with the \texttt{CDISORT} test case as we increase the number of layers from 40 to 140. For 140 layers, SH4 is within 2\% of \texttt{CDISORT} versus 9.7\% for SH2, illustrating that although twice as slow as SH2, SH4 is nearly five times more accurate when benchmarked against 16-stream \texttt{CDISORT}.

In cases where model accuracy is important and in the single scattering/asymmetry regions outlined in Section \ref{sec:heatmaps}, SH4 is the obvious choice due to its superior agreement with higher-fidelity models over its lower-term counter-model SH2. 
However, in the instance when rapid solutions are required, the additional computational expense of SH4 might be undesirable and the efficient SH2 will be the model of choice.

\section{Conclusion}
\label{sec:conclusion}
Following from the analysis conducted by \cite{rooney2023spherical}, we extended the spherical harmonics approach to solving the radiative transfer equation, implemented in modelling software \texttt{PICASO} \cite{natasha_batalha_2022_6419943}, to thermal emission.
In particular, we considered a four-term expansion of spherical harmonics, an increase from the original, two-stream implementation in \texttt{PICASO}, which we denoted Toon89 to reflect its heritage from \cite{toon1989rapid}.
The general spherical harmonics methodology for reflected light and thermal emission is largely the same, except for the source function, boundary conditions, and use cases. 
The main objective of this work was to build on the rigorous derivation of the model for reflected light studied by \cite{rooney2023spherical}, and explain the differences in the model for thermal emission.
Without re-deriving every equation in the model, we highlighted the differences incurred by considering a thermal source, and outlined the relevant matrix systems being solved by the model.

To explore the accuracy performance of the four-stream spherical harmonics model, we compared our results to \texttt{CDISORT} \citep{stamnes2000disort}.
When considered alongside two-term spherical harmonics and the two-stream Toon89 method, this analysis elucidated the increased efficacy of higher-order approximations in radiative transfer calculations for thermal emission, and also demonstrated the impact of the choice of phase function on the resulting spectra.
We studied the thermal spectra obtained via the two-stream Toon89 implementation, two and four-term spherical harmonics and \texttt{CDISORT} in Section \ref{sec:sample_atmos} for two different sample atmospheres.
This investigation highlighted that the choice of phase function has a large impact on the resultant spectra. The use of hemispheric mean in Toon89 created spectra that were largely different (up to 60\%) than those computed from SH2, which utilizes a Henyey-Greenstein phase function.
Additionally, we find that accuracy of the order of approximation (two versus four term) is highly dependent on the single scattering albedo and asymmetry of the cloud profile.
This motivated a deeper exploration of how the accuracy of the radiative transfer method depends on both values.

Therefore, we created a grid of models with a fixed atmosphere profile and varied asymmetry parameters and single-scattering albedos to study the Toon89 and SH4 models performance when compared to 32-stream \texttt{CDISORT}.
We found that Toon89 performs particularly well for the limiting cases of single-scattering albedo, namely $w_0=0$ and $w_0=1$, however suffered from substantial errors of around 60\% for high asymmetry and moderate single-scattering.
SH4 experiences maximum error of around -6\% for high single-scattering. 

Finally, we analyzed the timing-accuracy trade-off for the spherical harmonics methods when increasing the number of model layers. 
By calculating the maximum percentage difference between the thermal spectra produced by SH2 and SH4 with 16-stream, 140-layer \texttt{CDISORT}, we discussed the sacrifice in computational speed for model agreement.
This study elucidated that, although increasing model approximation order from two to four terms results in an increase in computational expense, the increase in accuracy when bench-marked against \texttt{CDISORT} is significant. 
The SH4 model took twice as long as SH2 to calculate the thermal spectra, but produced a result that was nearly five times more accurate when compared to 16-stream \texttt{CDISORT}, with a maximum percentage error of 2\%.
This analysis demonstrates that a sacrifice of computational expense might be acceptable when a significant increase in accuracy is required from the observational data accuracy, but may not be necessary if numerical efficiency is the priority.

In conclusion, we have demonstrated that increasing the order of approximation from two to four streams can produce significant improvement on model accuracy when compared with high-order \texttt{CDISORT}. 
The spherical harmonics analysis outlined in this paper is implemented in the \texttt{PICASO} framework, alongside the Toon89 methodology, and is available for download and use \cite{natasha_batalha_2022_6419943}. The Jupyter notebook, which reproduces our results, can be found on Github as well \href{https://github.com/natashabatalha/picaso/blob/9d4cbd672a75c1faf5297c3f1d74074018cd7ef3/docs/notebooks/10c_AnalyzingApproximationsThermal.ipynb}{\faCode} .

\begin{acknowledgements}
C.R.’s research was supported by an appointment to the NASA Postdoctoral Program at the NASA Ames Research Center, administered by Universities Space Research Association under contract with NASA. N.B. \& C.R. both acknowledge support from the NASA Astrophysics Division. Additionally, N.B. acknowledges support from NASA’S Interdisciplinary Consortia for Astrobiology Research (NNH19ZDA001N-ICAR) under award number 19-ICAR19\_2-0041. We thank Jeff Cuzzi and Sanford Davis for enlightening discussions about some of the finer points of radiative transfer in higher order approximations. Lastly we thank Arve Kylling for helpful discussions regarding \texttt{CDISORT}'s radiative transfer methodology. 
\end{acknowledgements}

\software{numba \citep{numba}, pandas \citep{mckinney2010data}, bokeh \citep{bokeh}, NumPy \citep{2020SciPy-NMeth}, \citep{walt2011numpy}, IPython \citep{perez2007ipython}, Jupyter, \citep{kluyver2016jupyter}, VIRGA \citep{natasha_batalha_2020_3759888,rooney2022new}, \texttt{PICASO} \citep{natasha_batalha_2022_6419943}, MATLAB \citep{MATLAB:2010}, A version of \text{PICASO} corresponding to these hyperlinks and the software used in this work is archived on Zenodo as v3.1 with DOI: 10.5281/zenodo.7765171}

\clearpage
\appendix
\section{Derivation of boundary condition}
\label{app:f_BC}
To derive the boundary condition for the upward flux $F^+(\tau_N)$ for an atmosphere that continues below the lower-most level in our model, we consider the intensity of emission at the surface given by \cite{mihalas1978stellar}, namely
\begin{equation}
		I(\tau_N,\mu) = B(\tau_N) + \mu\frac{\mathrm{d}B}{\mathrm{d}\tau}(\tau_N).
  \label{eq:lower_BC_mihalas}
\end{equation}
Recalling the expressions for $F^{\pm}(\tau) = 2\pi\int_0^{\pm1} I(\tau,\mu)\mu\mathrm{d}\mu$ and $f^{\pm}(\tau) = \pi\int_0^{\pm1} I(\tau,\mu)(5\mu^3-3\mu)\mathrm{d}\mu$, we obtain the boundary conditions
\begin{align}
	F^+(\tau_N) &= \pi\left(B(\tau_N) + \frac{2}{3}\frac{\partial B}{\partial\tau}(\tau_N)\right) ,\\
    f^+(\tau_N) &= -\frac{\pi B(\tau_N)}{4}.
\end{align}

Similarly, for the case of a hard-surface at the lower-most layer in our model, the intensity of emission at the surface is taken as that of a blackbody:
\begin{equation}
		I(\tau_N,\mu) = B(\tau_N).
\end{equation}
Proceeding as above we arrive at
\begin{align}
	F^+(\tau_N) &= \pi B(\tau_N),\\
    f^+(\tau_N) &= -\frac{\pi B(\tau_N)}{4}.
\end{align}
Including the effect of surface reflectivity $A_S$, hence, we obtain the boundary conditions
\begin{align}
	F^+(\tau_N) &= \pi B(\tau_N)+A_SF^-(\tau_N) ,\\
f^+(\tau_N) &= -\frac{\pi B(\tau_N)}{4} + A_Sf^-(\tau_N).
\end{align}

 Note that to minimize the effect of the choice of the lower boundary condition on the computed emergent flux it is always best practice to have the lowermost model layer lie at high optical depth at all wavelengths where practicable.

\clearpage
\bibliography{main}{}
\bibliographystyle{aasjournal}

\end{document}